\documentclass[%
 reprint,
 amsmath,amssymb,
 aps,
]{revtex4-2}

\usepackage{algpseudocode}
\usepackage[linesnumbered,ruled,vlined]{algorithm2e}
\usepackage{amsmath}     

\usepackage{amssymb}     
\usepackage{graphicx}    
\usepackage{tikz}

\usepackage{xcolor}
\usepackage{caption}
\usepackage{subcaption}
\usepackage{colortbl}

\usepackage{booktabs}
\usepackage{multirow}
\usepackage{siunitx}
\usepackage{adjustbox}

\usepackage[T1]{fontenc}
\usepackage[utf8]{inputenc}

\usepackage{hyperref}
\usepackage[mathlines]{lineno}

\begin{document}

\preprint{APS/123-QED}

\title{Learning-Optimized Qubit Mapping and Reuse to Minimize Inter-Core Communication in Modular Quantum Architectures}


\author{Sokea Sang}
\author{Leanghok Hour}
\author{Youngsun Han}\email{youngsun@pknu.ac.kr}
\affiliation{Department of AI Convergence, Pukyong National University, 45 Yongso-ro, Nam-gu, 48513, Busan, South Korea}

\date{\today}

\begin{abstract}
Modular quantum architectures have emerged as a promising approach for scaling quantum computing systems by connecting multiple Quantum Processing Units (QPUs). 
However, this approach introduces significant challenges due to costly inter-core operations between chips and quantum state transfers, which contribute to noise and quantum decoherence. 
This paper presents QARMA, a novel Qubit mapping approach using Attention-based deep Reinforcement learning (DRL) for Modular quantum Architectures, along with its extension, QARMA-R, which incorporates dynamic qubit reuse. 
Our approach combines an attention-based mechanism with Graph Neural Networks (GNN) to learn optimal qubit allocation, routing, and reuse strategies that minimize inter-core communications.
%
%
Our experimental results show significant improvements over state-of-the-art approaches. Compared to highly-optimized Qiskit with modular architecture configuration, QARMA-R reduces inter-core communications by up to 100\% (on average 86\%), while QARMA maintains 15-40\% improvement for larger circuits without reuse. Against traditional modular qubit mapping, our approach achieves 97-100\% reduction in inter-core operation.
The proposed methods advance quantum circuit compilation and enable the execution of larger quantum algorithms on resource-constrained modular quantum systems, thereby contributing to the growing body of research on scalable quantum computing architectures.
\end{abstract}

\maketitle


\section{Introduction}

\begin{figure*}[t]
\centering
\includegraphics[width=1\textwidth]{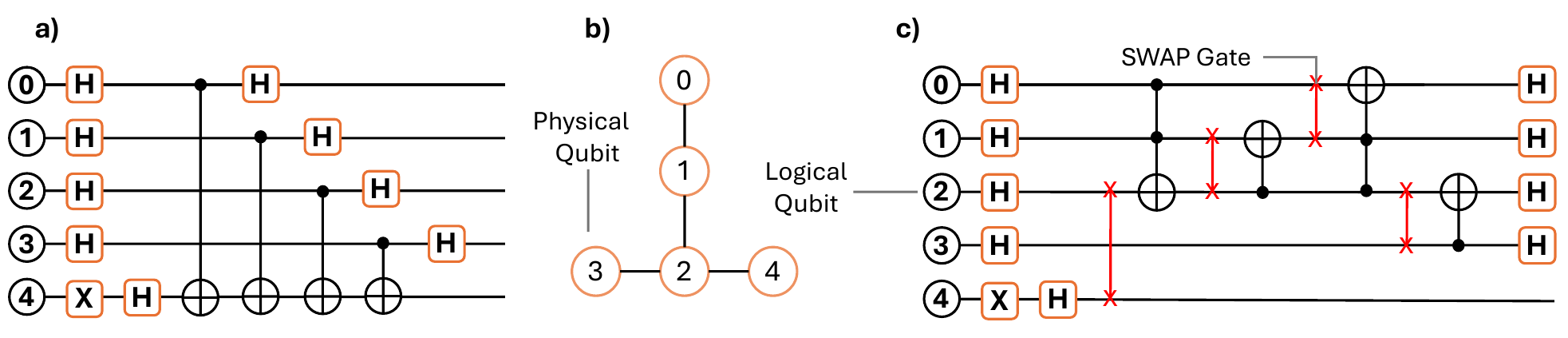}
\caption{Quantum circuit compilation process for the Bernstein-Vazirani (BV) algorithm~\cite{bv_alg} quantum circuit.
a) Original logical circuit with five qubits, showing Hadamard (H) gates and a CNOT operation.
b) Physical architecture with limited connectivity between qubits.
c) Compiled circuit with inserted SWAP gates (red) to accommodate hardware connectivity constraints.}
\label{circuit_compilation_fig}
\end{figure*}

Quantum computing promises exponential speedups for specific computational tasks that remain intractable in classical computers~\cite{nielsen2010quantum, shor1999polynomial, grover1996fast}. 
This field has witnessed remarkable progress in recent years, with quantum systems reaching sizes of more than 100 qubits~\cite{abughanem2024ibm}. 
However, to solve practically relevant problems, significantly larger quantum systems with thousands to millions of qubits will be needed~\cite{preskill2018quantum, arute2019quantum}. 
Scaling current noisy intermediate-scale quantum (NISQ) devices to these sizes presents substantial challenges because of physical constraints, control electronics limitations, and increased error rates~\cite{murali2019noise}.

Modular quantum architectures, which distribute quantum processing across multiple smaller interconnected cores, have emerged as a promising approach for overcoming these scaling challenges~\cite{monroe2014large, jnane2022multicore}. 
This approach, inspired by classical distributed computing paradigms, enables more manageable manufacturing, testing, and control of quantum hardware. 
However, it introduces a critical challenge: Quantum state transfers between cores are significantly more costly than operations within cores in terms of both execution time and error rates~\cite{rodrigo2021modelling, cuomo2023optimized}.

Therefore, efficiently allocating qubits across a modular quantum architecture is crucial for minimizing inter-core communication while respecting architectural constraints. 
This allocation problem is inherently an NP-hard optimization task~\cite{botea2018complexity, nannicini2022optimal} that extends the already challenging qubit mapping problem for single-core devices~\cite{li2019tackling}. 
Although several heuristic approaches have been proposed for modular architectures~\cite{baker2020time, bandic2023mapping, escofet2023hungarian}, they generally do not fully exploit the structure of quantum circuits or adapt well to different architectural topologies.
Furthermore, recent hardware advancements have introduced mid-circuit measurement and reset capabilities~\cite{corcoles2021exploiting, hua2022exploiting} that enable qubit reuse during computation. 
This functionality can reduce qubit requirements and significantly minimize inter-core communication; however, it has not been thoroughly integrated into qubit allocation strategies for modular architectures.

In this paper, we propose QARMA, a novel approach to the modular qubit allocation problem that leverages DRL with attention mechanisms~\cite{vaswani2017attention} and GNNs~\cite{gilmer2017neural}.
We also introduce QARMA-R, an extension that incorporates qubit reuse~\cite{hua2022exploiting}.
These approaches learn efficient allocation strategies dynamically that minimize inter-core communication while exploiting opportunities for qubit reuse.
The key contributions of this study are as follows.

(1) We introduce a transformer-based encoder architecture with GNNs that captures both global circuit structure and local qubit interactions to inform allocation decisions.

(2) QARMA-R implements a dynamic qubit reuse mechanism, which exploits mid-circuit measurement and reset operations to reduce resource requirements dramatically and minimize inter-core communications.

(3) Allocation decisions are guided by an attention-based pointer mechanism that directly outputs the probability of matching logical qubits with physical cores, enabling efficient assignment.

(4) We comprehensively evaluate realistic quantum circuit benchmarks, demonstrating significant reductions in inter-core communication. Compared to highly-optimized Qiskit, our approach achieves up to 100\% (on average 86\%) reduction with reuse enabled and maintains 15-40\% improvement even without reuse for larger circuits. Furthermore, compared to the QUBO-based mapping approach~\cite{bandic2023mapping}, QARMA-R achieves a 98-100\% reduction in inter-core connections or operations with reuse, and QARMA achieves a 97-100\% improvement without reuse.

Our approach builds on recent advances in applying reinforcement learning to combinatorial optimization problems~\cite{bello2016neural, kool2018attention, chen2019learning, berto2023rl4co}, adapting these techniques to the unique constraints and requirements of quantum circuit compilation. 
By formulating qubit allocation as a sequential decision-making process, our approach learns to allocate qubits in a way that minimizes inter-core communication by leveraging qubit reuse.

The remainder of this paper is organized as follows. 
Section~\ref{sec:background} provides the necessary background, including related work in Section~\ref{sec:related_work}. 
Section~\ref{sec:methodology} describes QARMA for qubit allocation and its extension QARMA-R with reuse capabilities. 
Section~\ref{sec:perf_eval} presents the experimental setup and results, followed by a discussion in Section~\ref{sec:discuss}. 
Section~\ref{sec:conclusion} concludes the study with a summary of the contributions and directions for future work. 

\begin{figure*}[t]
\centering
\includegraphics[width=1\textwidth]{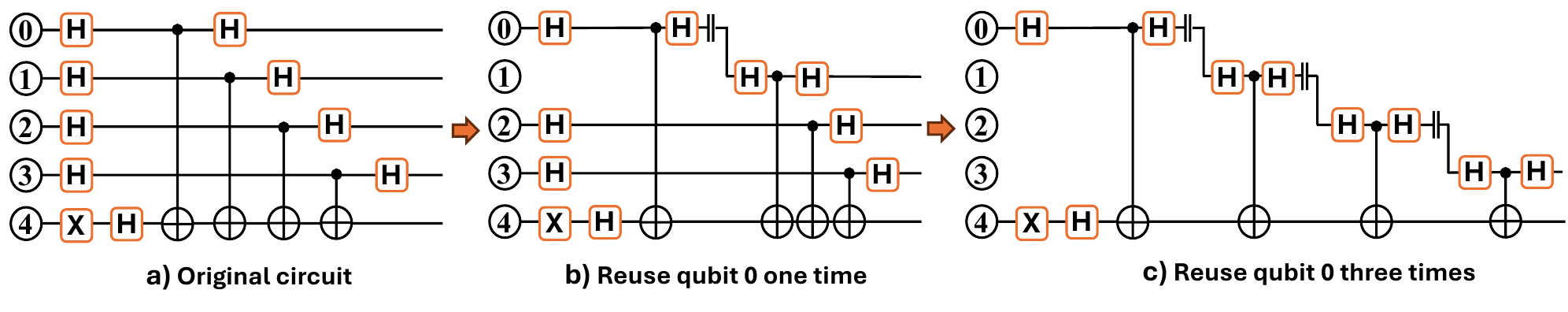}
\caption{Using Dynamic Circuit Support for the Bernstein-Vazirani (BV) algorithm~\cite{bv_alg} quantum circuit to reduce qubit usage: a) original five-qubit circuit; b) qubit 0 reused once, reducing to four qubits; c) qubit 0 reused three times, requiring only two qubits. Vertical double lines indicate measurement/reset operations.}
\label{qubit_reuse_fig}
\end{figure*}

\section{Background} \label{sec:background}
In this section, we introduce the key concepts and prior work necessary to understand our approach to qubit allocation in modular quantum architectures.

\subsection{Quantum Circuit Compilation} \label{subsec:quantum_circuit_compilation}

In the gate-based model of quantum computing, quantum programs are expressed as circuits consisting of a sequence of gates $G$ operating on logical qubits. 
Quantum compilation transforms these logical circuits into executable operations on physical quantum hardware~\cite{nannicini2022optimal, li2019tackling}.

Fig.~\ref{circuit_compilation_fig} illustrates the quantum circuit compilation. The original logical circuit (Fig.~\ref{circuit_compilation_fig}a) contains operations that assume that any qubit can interact with any other qubit. 
However, physical hardware (Fig.~\ref{circuit_compilation_fig}b) has limited connectivity, requiring the insertion of SWAP operations (Fig.~\ref{circuit_compilation_fig}c) to move quantum states between qubits and enable the execution of two-qubit gates between logical qubits that aren't adjacent in the physical architecture.

This transformation must address several key constraints: (1) physical qubits have limited connectivity, that is, two-qubit gates can only be executed between physically connected qubits; (2) the target hardware supports only a limited set of native gates; and (3) physical qubits and operations have varying error rates~\cite{murali2019noise}. 
The compilation process typically involves several steps, including gate decomposition, qubit mapping, scheduling, and other optimizations.~\cite{zhang2021time}.

The qubit mapping problem, which involves assigning logical qubits to physical qubits, is particularly challenging. 
In single-core architectures with sparse connectivity, this mapping often requires the insertion of SWAP operations to move the quantum states between physical qubits~\cite{li2019tackling, siraichi2018qubit}. 
This problem has been proven to be NP-complete~\cite{botea2018complexity, siraichi2018qubit}, with various heuristic~\cite{li2019tackling, zulehner2018efficient} and exact~\cite{nannicini2022optimal} approaches.

\subsection{Mid-Circuit Measurement and Qubit Reuse} \label{subsec:qubit_reuse}
Recently, quantum hardware platforms have begun to support dynamic circuits with mid-circuit measurement capabilities~\cite{corcoles2021exploiting}. 
Unlike traditional quantum circuits, where measurements occur only at the end of computation, dynamic circuits allow for measuring qubits during execution and making subsequent operations conditional on the measurement outcomes. 
This enables qubit reuse when combined with qubit reset operations. 
A single physical qubit can be used for multiple logical qubits at different points during circuit execution ~\cite{hua2022exploiting}.

Fig.~\ref{qubit_reuse_fig} demonstrates how qubit reuse can dramatically reduce the resource requirements. 
The original circuit (Fig.~\ref{qubit_reuse_fig}a) requires five qubits. 
By introducing mid-circuit measurement and reset operations, qubit 0 can be reused once its operations are complete (Fig.~\ref{qubit_reuse_fig}b), thereby reducing the qubit count to 4. 
Using this approach, qubit 0 can be reused multiple times (Fig. ~\ref{qubit_reuse_fig}c), requiring only two qubits to execute the entire circuit. 
This technique is particularly valuable for circuits with sequential dependencies, where many qubits complete their operations before others begin.

Qubit reuse through mid-circuit measurement offers three key benefits: (1) reduced qubit usage, allowing more extensive algorithms to run on resource-constrained hardware; (2) fewer SWAP operations, because logical qubits can be allocated to physical qubits that are optimally positioned for their specific operations; and (3) improved circuit fidelity by avoiding error-prone qubits or operations~\cite{hua2022exploiting}.

However, effective qubit reuse requires careful analysis of the qubit dependencies within a circuit. 
A logical qubit can only be reused if all its operations are completed and its state does not affect future operations, except through explicit measurement outcomes~\cite{hua2022exploiting, decross2023qubit}.

\subsection{Circuit Slicing and Qubit Allocation}
\label{subsec:circuit_slicing}
A common approach to quantum circuit compilation for modular architectures involves \textit{circuit slicing}, which partitions a circuit into time slices containing gates that can be executed in parallel~\cite{baker2020time, russo2024attention}. 
For each slice, logical qubits must be allocated to physical cores such that: (1) qubits involved in the same gate are allocated to the same core ("friend qubits"); (2) the number of qubits allocated to each core does not exceed its capacity; and (3) inter-core communication between consecutive slices is minimized~\cite{russo2024attention}.

The process is typically iterative, mapping each slice in sequence and accounting for both the constraints of the current slice and the allocation of the previous slice. 
Various heuristics for this allocation have been proposed, including look-ahead weight assignment~\cite{baker2020time}, quadratic optimization formulations~\cite{bandic2023mapping}, and Hungarian algorithm-based approaches~\cite{escofet2023hungarian}.

\begin{figure*}[t]
\centering
\includegraphics[width=1\textwidth]{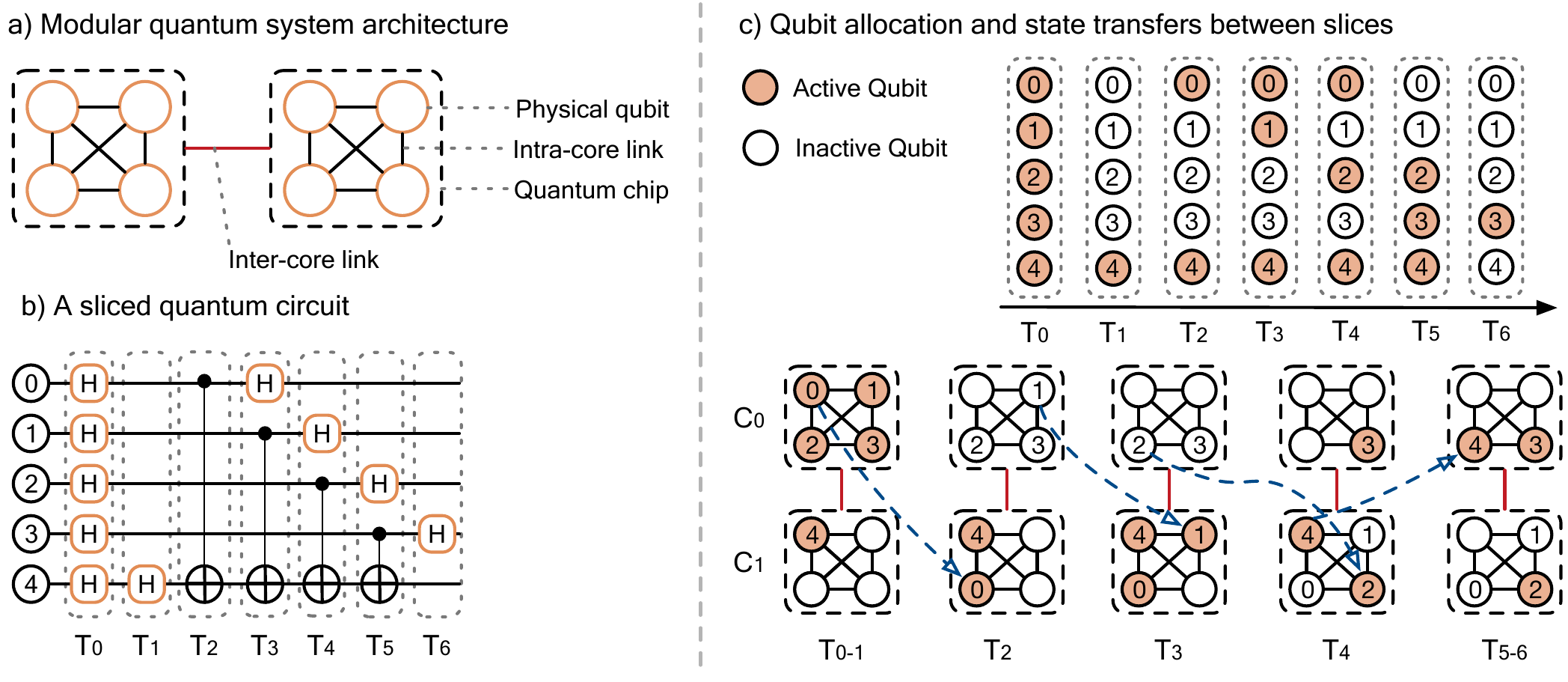}
\caption{Modular quantum architecture and circuit execution: a) A multi-core quantum architecture with two cores, each containing four fully-connected physical qubits, linked by inter-core connections; b) A sliced quantum circuit where each vertical region represents a time slice containing gates that can be executed in parallel; c) Qubit allocation and state transfers across time slices, showing how logical qubits (numbered 0-4) are mapped to physical cores ($c_0$ and $c_1$) at different time steps ($T_0$ through $T_6$). Arrows indicate quantum state transfers between cores, which are more costly than intra-core operations.}
\label{modular_architecture_fig}
\end{figure*}

Unlike single-core compilation, where SWAP gates are inserted to move quantum states between connected physical qubits, modular architectures rely on quantum state transfer protocols between cores, which are typically more expensive in terms of time and fidelity~\cite{rodrigo2021modelling}

\subsection{Modular Quantum Systems} \label{subsec:modular_quantum_system}

Scaling quantum computers to thousands or millions of qubits faces significant challenges with monolithic architectures, including manufacturing yield, control of electronics density, and qubit crosstalk~\cite{arute2019quantum, smith2022scaling}. 
Modular quantum architectures distribute quantum processing across multiple small cores (or modules) connected by quantum communication channels~\cite{monroe2014large, jnane2022multicore, russo2024attention}.

Fig.~\ref{modular_architecture_fig} illustrates the modular quantum architecture and execution of a quantum circuit across it. 
In Fig.~\ref{modular_architecture_fig}a, we observe a system with two quantum cores, each containing four fully connected physical qubits (represented by white circles), with inter-core links connecting the cores. 
Fig.~\ref{modular_architecture_fig}b shows a quantum circuit divided into time slices, where the operations within each slice can be executed in parallel.

Fig.~\ref{modular_architecture_fig}c demonstrates how logical qubits are allocated to physical cores over time and the resulting quantum state transfers. 
Initially, at time $T_0$, logical qubits 0-3 are assigned to core $c_0$ and logical qubit 4 to core $c_1$. 
As the circuit progresses through time steps $T_2$ to $T_n$, logical qubits must be transferred between cores (indicated by red arrows) to enable gate operations between qubits allocated to different cores. 
These inter-core quantum state transfers are significantly more time-consuming and error-prone than operations within a core.

In these architectures, operations within a core can be performed with relatively high fidelity, whereas operations between cores, particularly quantum state transfers, incur higher error rates and latency~\cite{rodrigo2021modelling}. 
The multi-core interconnect must support both classical communication for control and quantum communication for state transfer between cores~\cite{russo2024attention}.
Various physical implementations have been proposed, including photonic links for superconducting qubits~\cite{xue2022quantum} and ion shuttling for trapped-ion systems~\cite{pino2021demonstration}.

The key optimization objective in modular architectures is to minimize inter-core communications while respecting each core's capacity and connectivity constraints~\cite{baker2020time, bandic2023mapping}. 
This introduces a new dimension to the compilation problem beyond traditional single-core mapping.

\subsection{Reinforcement Learning for Combinatorial Optimization} \label{subsec:rl4co}
Reinforcement learning (RL) has emerged as a powerful approach for solving combinatorial optimization problems~\cite{bello2016neural, bengio2021machine}. 
In particular, attention-based neural architectures have shown promising results in routing problems~\cite{kool2018attention}, scheduling~\cite{chen2019learning}, and device placement~\cite{mirhoseini2017device}.
In the RL framework, a policy $\pi_\theta$ parameterized by $\theta$ learns to make sequential decisions that maximize the expected rewards. 
In combinatorial optimization problems, solutions are constructed step by step, with the reward typically based on the quality of the final solution.
Recent studies have applied RL to quantum circuit compilation tasks, primarily for single-core architectures~\cite{fosel2021quantum, zhang2021time}. 
These approaches demonstrate that RL can learn effective heuristics that adapt to different circuit structures and hardware topologies.
For large action spaces, attention mechanisms provide a powerful means of focusing on the relevant parts of the problem state~\cite{vaswani2017attention, kool2018attention}. 
Graph Neural Networks (GNNs) offer a natural way to represent structured data like quantum circuits, capturing the complex dependencies between qubits and operations~\cite{russo2024attention}.
In our study, we introduce QARMA and its extension QARMA-R, which address the unique challenges of qubit allocation in modular architectures. 
The base QARMA combines transformer encoders and GNNs to learn allocation policies that minimize inter-core communication, while QARMA-R extends this capability by integrating dynamic qubit reuse opportunities through mid-circuit measurement and reset operations.

\section{Related Work} \label{sec:related_work}
Several approaches have been studied for qubit mapping in modular architectures. 
Here, we discuss relevant methods based on QUBO, comprehensive toolchains like Qiskit, and other RL techniques, highlighting how our work differs.

Bandic et al.~\cite{bandic2023mapping} proposed formulating the qubit allocation problem as a Quadratic Unconstrained Binary Optimization (QUBO) problem, which directly encodes mapping constraints and optimization objectives into a quadratic function. 
This approach comprehensively encompasses the solution landscape without relying on limited look-ahead functions as in ~\cite{baker2020time}, decouples optimization from the objective function, potentially leverages quantum annealers, and naturally expresses the $k$-partitioning problem often used in modular systems. 
The QUBO approach splits circuits into slices, formulates qubit assignment as a graph-partitioning problem, and penalizes inter-core communication through quadratic terms. 
While QUBO provides a comprehensive formulation, it cannot exploit dynamic qubit reuse opportunities during circuit compilation, a key capability of our QARMA-R.

IBM's Qiskit~\cite{Qiskit2019} offers extensive optimization capabilities through its transpiler, which transforms high-level quantum circuits into executable operations while optimizing for performance metrics. 
The transpilation pipeline in Qiskit~\cite{QiskitTranspiler2021, QiskitTranspiler} includes analysis passes, transformation passes, layout selection, routing, and post-mapping optimizations. 
At its highest optimization level, Qiskit employs methods such as the SABRE algorithm~\cite{li2019tackling} for qubit mapping and the Dense Layout algorithm~\cite{murali2019noise} for layout selection. 
However, Qiskit has several limitations for modular architectures: it's primarily designed for monolithic processors, doesn't systematically exploit qubit reuse opportunities despite supporting dynamic circuits~\cite{QiskitDynamicCircuits}, employs heuristics that don't adapt to specific circuit structures, and uses a cost model that doesn't fully capture the penalties of inter-module state transfers.

Russo et al.~\cite{russo2024attention} recently proposed an attention-based deep reinforcement learning approach for qubit allocation that shares conceptual similarities with our methodology. 
While a direct empirical comparison was not possible as their implementation is not publicly available and the training datasets differ significantly, we can identify several key conceptual differences: (1) we incorporate mid-circuit measurement and reset operations enabling physical qubit reuse, while they employ a static one-to-one mapping; (2) our approach optimizes both for reduced communication and qubit count through reuse, creating dramatically different allocation strategies; (3) we extend the action space to include decisions about measurement and reset points; and (4) our method demonstrates superior performance over Qiskit and QUBO approaches, often achieving zero inter-core communications with reuse enabled, while Russo et al. show improvements over black-box optimizers but report performance degradation on structured circuits like QFT.
While both approaches demonstrate significant improvements over classical methods, our approach's incorporation of dynamic qubit reuse represents a fundamental advancement addressing both communication and resource constraints critical for practical modular quantum computing. We discuss in more detailed of indirect comparison with this approach in Section~\ref{sec:discuss}.

In addition to the methods discussed, other advanced compilation frameworks have recently been proposed. For instance, the DRL-based DEAR framework by Huang et al.~\cite{dear-framework} leverages reinforcement learning to find an optimal initial mapping and uses A* search for routing on monolithic architectures. While it shares the use of DRL, its core focus is on single-processor qubit mapping and SWAP insertion. It does not address the distinct challenges of modular systems, such as circuit partitioning, qubit allocation across multiple cores, and the explicit minimization of costly inter-chip communication, which are the primary optimization targets of QARMA. Therefore, a direct experimental comparison is not applicable as the two frameworks are designed to solve different optimization problems.

Similarly, the TITAN framework ~\cite{titan-framework} presents a co-design of a novel photonic interconnection hardware for distributed trapped-ion quantum computers and a corresponding compiler. Its mapping strategy is based on a classical, heuristic-based hierarchical partitioning algorithm (Kernighan-Lin) tailored specifically to its unique hardware. In contrast, QARMA is a hardware-agnostic DRL-based approach designed to learn generalizable mapping policies for any modular architecture defined by a topology graph. The fundamental differences between the target hardware platforms (trapped-ion vs. general modular) and the algorithmic approaches (heuristic vs. DRL), combined with the unavailability of a public implementation of TITAN, preclude a direct and meaningful performance comparison. Our work instead focuses on comparing QARMA with baseline methods applicable to general modular architectures, such as highly optimized Qiskit and QUBO-based mappers.

\begin{figure*}[t]
    \centering
    \includegraphics[width=\textwidth]{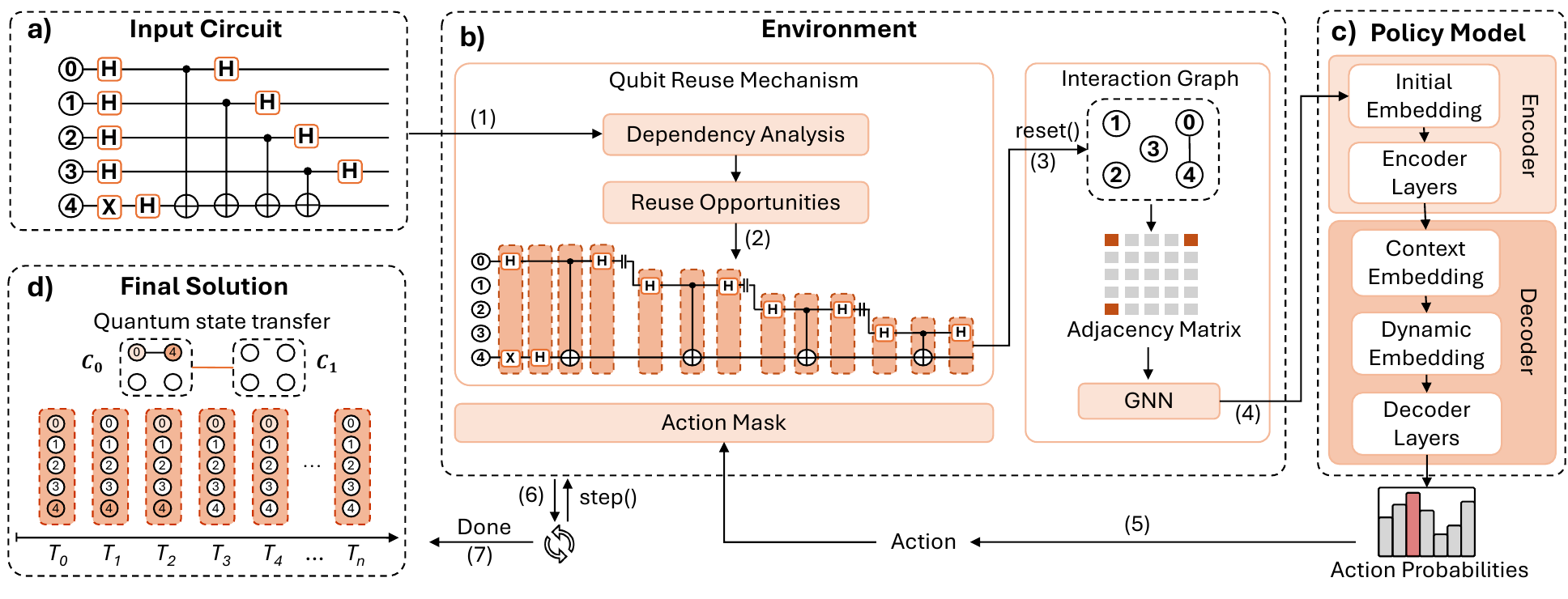}
    \caption{Overall of our approach for qubit allocation with reuse in modular quantum architectures. a) Input quantum circuit with multiple qubits and gates; b) environment that performs dependency analysis to identify qubit reuse opportunities and generate action masks; c) an attention-based policy model with encoder-decoder architecture that processes the circuit's interaction graph and produces allocation decisions; d) final allocation solution showing quantum state transfers between cores across time steps. The numbered arrows (1-7) indicate the workflow sequence through our approach.}
    \label{overall_QARMA_and_QARMA-R_fig}
\end{figure*}

\section{Proposed Methodology} \label{sec:methodology}
This section presents QARMA and QARMA-R for qubit allocation in modular quantum architectures. We provide an overview of our approach, including the qubit reuse mechanism, followed by the transformer-based circuit architecture and training process for minimizing inter-core communication.

\subsection{Overall Architecture}
\label{subsec:overall_DRL}

Illustrated in Fig.~\ref{overall_QARMA_and_QARMA-R_fig}, our approach comprises four main components: an input quantum circuit, an environment with a qubit reuse mechanism, an attention-based policy model, and a final allocation solution. The workflow begins with a sequential process, indicated by numbered arrows (1–7), as detailed below.

\subsubsection{Problem Representation and Environment Processing}

The workflow begins with the input quantum circuit shown in Fig.~\ref{overall_QARMA_and_QARMA-R_fig}a, which consists of five logical qubits (labeled 0-4) with Hadamard (H) gates, an X gate, and multiple CNOT operations, depicted as vertical connections. 
This circuit serves as the initial representation of the problem. 
In step (1), the input circuit is passed to the Qubit Reuse Mechanism within the environmental component (Fig.~\ref{overall_QARMA_and_QARMA-R_fig}b). 
The environment plays a central role in our approach and contains three critical subcomponents:

\begin{itemize}
    \item \textbf{Dependency Analysis:} This module analyzes the lifetime of each qubit throughout the circuit to identify potential reuse opportunities. It creates a dependency graph that tracks when qubits are created, used in operations, and made available for potential resets (details in Section~\ref{subsec:qubit_reuse_impl}).
    
    \item \textbf{Reuse Opportunities:} Based on the dependency analysis, this module identifies points where qubits can be measured and reset, effectively allowing their physical resources to be reused for other logical qubits.
    
    \item \textbf{Action Mask:} This component constrains the action space to maintain solution validity by generating masks that prevent illegal allocations, such as exceeding core capacity or violating connectivity constraints (see in detail Appendix~\ref{app:action_mask}).
\end{itemize}

In step (2), the Qubit Reuse Mechanism transforms the original circuit by inserting measurement and reset operations at appropriate points, as shown in the lower portion of Fig.~\ref{overall_QARMA_and_QARMA-R_fig}b. 
The circuit is then decomposed into time slices (vertical regions), where each slice contains gates that can be executed in parallel. 
This slicing is crucial for the subsequent allocation process because it defines the temporal structure of execution.

\subsubsection{Policy Model Architecture}

Step (3) represents the generation of the problem instance from the environment using the \texttt{reset()} function in the Policy Model Architecture (details in Section~\ref{subsec:neural_arch} and Section~\ref{subsec:transformer_encoder}) shown in Fig.~\ref{overall_QARMA_and_QARMA-R_fig}c. 
This instance includes a sliced circuit and interaction graph representation in step (3), where nodes represent qubits and edges represent gates between qubits. 
This graph is represented as an adjacency matrix that captures the connectivity patterns among qubits, providing crucial structural information for allocation decisions.

In Fig.~\ref{overall_QARMA_and_QARMA-R_fig}c, the model consists of two primary components: an encoder and a decoder. 
In step (4), the initial embedding module processes the circuit structure using GNNs~\cite{gilmer2017neural} to capture the qubit interaction patterns. 
The encoder layers then apply self-attention mechanisms~\cite{vaswani2017attention} to incorporate global information across the time slices. 
During decoding, the context embedding combines circuit-level, slice-level, and qubit-level information to create a comprehensive context vector. 
In contrast, the dynamic embedding generates core-specific representations that account for current capacity constraints and communication costs.

\subsubsection{Action Generation and Execution}

The decoder produces action probabilities in step (5), determining the optimal core assignment for each qubit. 
These action probabilities are filtered through action masks that prevent invalid allocations, ensuring that the core capacity constraints and friend qubit requirements are satisfied. 
The selected action is then executed in the environment through the \texttt{step()} function in step (6), which updates the allocation state and computes the reward based on the inter-core communication costs.

\subsubsection{Solution Generation and Improvement}

Finally, this process is repeated until all qubits across all the time slices are allocated (step 7), resulting in a complete allocation solution in step (7); the iterative process produces a complete allocation solution, as shown in Fig.~\ref{overall_QARMA_and_QARMA-R_fig}d. 
The Final Solution component evaluates the allocation quality by measuring the quantum state transfer requirements between cores.

The upper portion of Fig.~\ref{overall_QARMA_and_QARMA-R_fig}d illustrates how qubit reuse enables efficient allocation: logical qubit 0 (in core $c_0$) can be measured, and its physical resources reused, eliminating the need for an inter-core transfer to core $c_1$.

The lower timeline shows the allocation of logical qubits (0-4) across time steps ($T_0$–$T_n$). The effectiveness of our approach is demonstrated by the minimal inter-core communication required, as the policy has learned to exploit qubit reuse opportunities. 
In the optimal case shown, by correctly identifying and exploiting qubit reuse opportunities, the model can allocate circuits to multiple cores with minimal or no inter-core connections, thereby significantly reducing communication overhead while maintaining quantum coherence.


\subsection{Sequential Decision-Making Process}

\begin{algorithm}[h!]
\caption{DRL-based Qubit Allocation}
\label{alg:qubit-allocation}
\SetKwInOut{Input}{Input}
\SetKwInOut{Output}{Output}
\Input{
    circuit with $T$ time slices and $Q$ logical qubits\;
}
\Input{
    topology with $C$ cores and distance matrix\;
}
\Input{
    core\_capacities: physical qubit capacity\;
}
\Input{
    policy\_network: trained attention-based\;
}
\Output{allocation: 3D tensor mapping qubits to cores across time slices}
\BlankLine
allocation $\gets$ Initialize3DTensor($T$, $Q$, $C$, 0)\;
core\_usage $\gets$ Initialize2DTensor($T$, $C$, 0)\;
reset\_indicators $\gets$ IdentifyReusableQubits(circuit)\;
total\_reward $\gets$ 0\;
\For{$t \gets 1$ \KwTo $T$}{
    state $\gets$ UpdateEnvironmentState($t$, allocation, core\_usage)\;
    qubit\_order $\gets$ DetermineProcessingOrder(circuit[$t$], state)\;
    \For{each $q$ in qubit\_order}{
        encoded\_state $\gets$ EncodeState(state, circuit, $q$, $t$)\;
        core\_probs $\gets$ policy\_network.forward(encoded\_state)\;
        selected\_core $\gets$ SampleFromDistribution(core\_probs)\;
        allocation[$t$, $q$, selected\_core] $\gets$ 1\;
        \If{reset\_indicators[$t$, $q$] = 0}{
            core\_usage[$t$, selected\_core] += 1\;
        }
    }
    slice\_reward $\gets$ 0\;
    \If{$t > 1$}{
        \For{$q \gets 1$ \KwTo $Q$}{
            prev\_core $\gets$ GetAllocatedCore(allocation, $t-1$, $q$)\;
            curr\_core $\gets$ GetAllocatedCore(allocation, $t$, $q$)\;
            \If{prev\_core $\neq$ curr\_core}{
                comm\_cost $\gets$ architecture[prev\_core, curr\_core]\;
                slice\_reward -= comm\_cost \Comment{Negative cost as reward}
            }
        }
    }
    total\_reward += slice\_reward\;
}
\For{$t \gets 1$ \KwTo $T$}{
    \For{$q \gets 1$ \KwTo $Q$}{
        \If{IsQubitCompleted($q$, $t$, circuit)}{
            reset\_indicators[$t$, $q$] $\gets$ 1 \; \tcp{Mark qubit as reusable}
        }
    }
}
\KwRet{allocation}\; 
\end{algorithm}

We formulate the qubit allocation problem as a sequential decision-making process in which logical qubits are allocated to physical cores individually. This approach enables our agent to consider the impact of each allocation decision on future allocation.

Algorithm~\ref{alg:qubit-allocation} details the sequential decision-making process with the following key components.

\paragraph{Inputs and Initialization.} The algorithm uses four key inputs: (1) a quantum circuit described as a sequence of time slices containing gate operations, (2) the architecture topology specifying distances between cores, (3) the capacity constraints of each core, and (4) a trained policy network. 
The process begins with initializing data structures to track allocations, core resource usage, qubit reuse opportunities, and total rewards (lines 1-4).

\paragraph{Slice-by-Slice Processing}
For each time slice (line 5), we update the environment state (line 6) and determine an optimized processing order for qubits based on their connectivity patterns and reuse potential (line 7). Our implementation uses a heuristic to establish this order, specifically by prioritizing qubits involved in the most two-qubit gates within the current slice. By addressing the most constrained allocation decisions first, this approach helps to prune the search space for subsequent qubit assignments. This ordering is crucial for satisfying the dependency constraints while maximizing allocation efficiency.

\paragraph{Autoregressive Allocation}
The policy network makes allocation decisions sequentially (lines 8-15), with each decision conditioned on previous allocations. 
This autoregressive approach enables the model to adapt its strategy based on partial allocations, significantly outperforming the traditional heuristics. 
Core capacity constraints are enforced by tracking resource usage, with special handling of reusable qubits that do not consume additional physical resources.

\paragraph{Communication Cost Evaluation.} A distinguishing feature of our approach is explicitly modeling inter-core communication costs (lines 16-23). 
For each qubit that changes cores between consecutive time slices, we compute a communication penalty based on the topology matrix of the architecture. 
This mechanism directly encourages the policy to minimize quantum state transfers while respecting logical dependencies.

\paragraph{Qubit Reuse Identification.} In the final phase (lines 24-27), we identify the qubits that have completed all their operations in the circuit and mark them as candidates for reuse. 
This enables efficient resource utilization by recycling physical qubits for different logical qubits across time slices, thereby dramatically reducing the total physical qubit requirement.

\paragraph{Output.} The algorithm returns a 3D tensor representing the allocation of each logical qubit to a physical core across all the time slices (line 28).

By embedding architectural constraints and qubit reuse opportunities directly into the reinforcement learning framework, our approach learns allocation strategies that significantly outperform conventional heuristics regarding resource efficiency and communication cost minimization.

\paragraph{Relationship between Logical Qubits, and Physical Cores}
QARMA and QARMA-R minimize communication costs between physical cores while respecting the logical circuit structure through the following key relationships:

\begin{enumerate}
\item \textbf{Logical-to-Physical Mapping:} Decision variable $x_{t,q,c}=1$ when logical qubit $q$ is assigned to core $c$ at time slice $t$, with interacting qubits assigned to the same core.
\item \textbf{Core Capacity Constraints:} Each core $c$ has capacity $P_c$, limiting its qubit assignments.
\item \textbf{Communication Costs:} When qubit $q$ moves between cores ($x_{t,q,c_1}=1$, $x_{t+1,q,c_2}=1$ where $c_1 \neq c_2$), it incurs cost $D_{c_1,c_2}$ based on architecture topology.
\item \textbf{Qubit Reuse:} In QARMA-R, when qubit $q_1$ completes operations at time $t$, it can be reused for qubit $q_2$ 
\end{enumerate}

Our frameworks learn allocation strategies by maximizing a reward function that minimizes total communication cost:
\begin{equation}
R = -\sum_{t=1}^{T-1} \sum_{q=1}^{Q} \sum_{c_1=1}^{C} \sum_{c_2=1}^{C} x_{t,q,c_1} \cdot x_{t+1,q,c_2} \cdot D_{c_1,c_2}
\end{equation}
The negative sign converts cost minimization to reward maximization for the RL agent.

\subsection{Qubit Reuse Mechanism Implementation}
\label{subsec:qubit_reuse_impl}

One of the key innovations in our approach is integrated the dynamic qubit reuse mechanism~\cite{hua2022exploiting}, which identifies and exploits opportunities for measurement and reset operations to reduce physical qubit requirements. 
Our implementation employs a dependency-driven algorithm to identify optimal reuse pairs while minimizing increases in circuit depth.

Given a quantum circuit, our `optimize\_qubit\_reuse` function analyzes qubit lifetimes to identify potential reuse candidates:

\begin{enumerate}
    \item We first compute the last operation index for each qubit in the circuit, recording the final timestep at which each qubit is used.
    \item We then compute the first operation index for each qubit, indicating when each qubit begins its operation.
    \item Potential reuse pairs $(q_1, q_2)$ are identified where the last operation of $q_1$ occurs before the first operation of $q_2$.
\end{enumerate}

For each candidate reuse pair, we evaluate the impact on circuit depth using a weighted cost function:
\begin{equation}\label{eq:qubit_reuse_weight}
\begin{split}
\textit{cost}(q_1, q_2) = & \, w_0 \cdot \Delta\textit{depth} + w_1 \cdot \textit{last\_idx}(q_1) \\
& + w_2 \cdot |\textit{last\_idx}(q_1) - \textit{first\_idx}(q_2)|
\end{split}
\end{equation}

where $w_0$, $w_1$, and $w_2$ are tunable weights balancing: (1) the increase in circuit depth due to measurement and reset operations; (2) preference for reusing qubits that complete early; and (3) preference for minimizing the gap between reuse operations. In this work, we set these to a default, equal weighting ($w_0 = w_1 = w_2 = 1$) to provide balanced consideration of all three objectives without specializing the reuse strategy towards any single factor. This neutral approach allows us to evaluate the core benefits of reuse while avoiding excessive increases in circuit depth. We acknowledge that further fine-tuning of these weights could yield additional improvements, representing an important direction for future optimization efforts, so we conduct a sensitivity analysis using five different weight configurations on a subset of complex benchmark circuits in Appendix~\ref{sec:sensitivity_analysis}.

The algorithm iteratively selects the best reuse pair, modifies the circuit to insert measurement and reset operations, and continues until no more beneficial reuse opportunities remain. The result is a modified circuit with significantly reduced qubit requirements, along with a mapping that tracks how logical qubits are chained together through reuse.

\subsection{Policy Model}
\label{subsec:neural_arch}

While Section \ref{subsec:transformer_encoder} described the transformer architecture, we now detail our custom policy model components specifically designed for the qubit allocation problem (Fig.~\ref{overall_QARMA_and_QARMA-R_fig}c):

\begin{enumerate}
    \item \textbf{Initial Embedding Module:} Our \textit{InitialEmbedding} component generates embeddings for the circuit structure and core topology. It processes circuit slices through a GNN to capture qubit interaction patterns and simultaneously encodes the core topology to represent the communication costs between cores.
    
    \item \textbf{Context Embedding Module:} The \textit{ContextEmbedding} component synthesizes information about the current decision state, integrating the circuit-level embeddings, the current time slice, and the qubit being allocated to provide comprehensive context for decision-making.
    
    \item \textbf{Dynamic Embedding Module:} The \textit{DynamicEmbedding} component is crucial for adapting to the changing state during allocation. It incorporates three dynamic features for each core:
    \begin{itemize}
        \item Current remaining capacity, which affects the feasibility of further allocations
        \item Distance from the previous allocation of the current qubit, representing the communication cost 
        \item Guidance information from hardware-specific hints (when available)
    \end{itemize}
    
    \item \textbf{Pointer Attention Mechanism:} Finally, our model uses an attention-based pointer mechanism that directly outputs the probability of allocating the current qubit to each core. This mechanism allows the model to focus on relevant cores based on the current context and dynamic state.
\end{enumerate}

This architecture enables effective representation learning across the hierarchical nature of the problem: circuit-level structure, slice-level interactions, and qubit-level decision-making. By factoring in both static circuit properties and dynamic allocation state, our model can adapt its strategy as allocation progresses.

\subsection{Transformer Circuit Encoder-Decoder}
\label{subsec:transformer_encoder} 

While GNNs capture local circuit structures \cite{gilmer2017neural}, we use a transformer architecture \cite{vaswani2017attention} to model long-range temporal dependencies across time slices, which is crucial for optimizing qubit reuse and minimizing inter-core communication globally.

The core is multi-head self-attention \cite{vaswani2017attention}. This enables the model to weigh the relevance of different time slices in allocation decisions. Input slice embeddings ($X = [x_1, ..., x_T]$) are transformed using the scaled dot-product attention mechanism as follows:
\begin{equation} \label{eq:attention}
\text{Attention}(Q, K, V) = \text{softmax}\left(\frac{Q \cdot K^T}{\sqrt{d_k}}\right)V
\end{equation}
Here, the query ($Q$), key ($K$), and value ($V$) matrices are the learned projections of the input $X$. The mechanism computes attention weights based on query–key compatibility and then scales the value vectors to produce output embeddings that incorporate context from all slices. For more details, see Vaswani et al. \cite{vaswani2017attention}.

We incorporate sinusoidal positional encodings \cite{vaswani2017attention} and relative position representations \cite{shaw2018self} to provide a temporal context. Our implementation uses multi-head attention ($h=8$ heads, model dimension $d=256$). The encoder and decoder stacks each consist of $N=3$ layers containing multi-head attention and feed-forward sublayers with residual connections and layer normalization \cite{vaswani2017attention}. The decoder employs masked self-attention to maintain autoregressive prediction for sequential allocation \cite{kool2018attention}.

This combination of GNNs and transformers provides comprehensive circuit understanding, enabling effective allocation and reuse strategies.

\subsection{Training Process}
We implemented QARMA and QARMA-R using the RL4CO (Reinforcement Learning for Combinatorial Optimization) library~\cite{berto2023rl4co}, which provides a unified and modular implementation of reinforcement learning algorithms for solving combinatorial optimization problems. 
This framework offers several advantages for our task, including flexible policy definitions, efficient environment implementation, and support for various training regimes.

We optimized the policy networks using the Adam optimizer~\cite{kingma2014adam} with an initial learning rate of $1.0 \times 10^{-4}$. 
Also, we employed a batch size of 512 circuit instances per training step to provide stable gradient updates. 
The training process consisted of 100 epochs. To ensure the policy generalizes well and does not overfit to a fixed dataset, we employed a dynamic data generation strategy. For each epoch, a new training batch of 500 quantum circuit instances was randomly generated on-the-fly, each with varying sizes and structures. 
We used separate sets of 100 circuit instances for validation and testing.

The reinforcement learning algorithm follows the REINFORCE paradigm with a specialized rollout baseline~\cite{kool2018attention}, which compares the current policy's performance with a greedy rollout to reduce the variance in gradient estimates. 
We implemented a modified baseline update strategy that accepts improvements only when the candidate policy outperforms the current baseline and the global minimum threshold. 
This approach helps maintain training stability while allowing for consistent improvement.

For each training instance, we dynamically analyzed the circuit structure to identify potential reuse opportunities by constructing a dependency graph that tracks when qubits complete their operations and become available for measurement and reset. 
This analysis informs the policy's decision-making process, enabling it to exploit qubit reuse opportunities when they are beneficial.

\begin{table}[t]
\centering
\caption{Benchmark circuits for the performance evaluation. The top section (white background) lists standard benchmark circuits, while the bottom section (gray background) lists randomly generated circuits categorized by scale (s-small, m-medium, l-large).}
\label{tab:combined-circuits}
\begin{tabular}{@{}l c c c@{}}
\toprule
\toprule
\textbf{Circuit Name} & \textbf{Qubits} & \textbf{Depth} & \textbf{Two-Qubit Gates} \\
\midrule
4mod5-bdd-287   & 16 & 41  & 31  \\
ex1-226         & 16 & 5   & 5   \\
graycode6-47    & 16 & 5   & 5   \\
xor5-254        & 16 & 5   & 5   \\
bv-n10          & 10 & 13  & 9   \\
cc-n10          & 10 & 11  & 9   \\
cnt3-5-179      & 16 & 61  & 85  \\
multiply-n13    & 13 & 42  & 43  \\
sym9-146        & 16 & 127 & 148 \\
rd53-311        & 16 & 124 & 124 \\
cnt3-5-180      & 16 & 209 & 215 \\
system9         & 16 & 127 & 150 \\
wim-266         & 16 & 514 & 427 \\
sym6-316        & 16 & 135 & 123 \\
rd84-142        & 16 & 110 & 154 \\
multiplier-n15  & 15 & 49  & 30  \\
ising-model-13  & 16 & 71  & 120 \\
ising-model-16  & 16 & 71  & 150 \\
cm152a-212      & 16 & 684 & 532 \\
square-root-n18 & 18 & 203 & 118 \\
multiplier-n45  & 45 & 462 & 306 \\
\midrule
\rowcolor[gray]{0.9} s-10q-16d-13cx  & 10 & 16 & 13 \\
\rowcolor[gray]{0.9} s-30q-5d-9cx    & 30 & 5  & 9  \\
\rowcolor[gray]{0.9} s-18q-10d-8cx   & 18 & 10 & 8  \\
\rowcolor[gray]{0.9} s-26q-9d-15cx   & 26 & 9  & 15 \\
\rowcolor[gray]{0.9} s-26q-9d-14cx   & 26 & 9  & 14 \\
\rowcolor[gray]{0.9} m-48q-7d-17cx   & 48 & 7  & 17 \\
\rowcolor[gray]{0.9} m-45q-8d-29cx   & 45 & 8  & 29 \\
\rowcolor[gray]{0.9} m-59q-11d-38cx  & 59 & 11 & 38 \\
\rowcolor[gray]{0.9} m-53q-18d-48cx  & 53 & 18 & 48 \\
\rowcolor[gray]{0.9} m-60q-15d-67cx  & 60 & 15 & 67 \\
\rowcolor[gray]{0.9} l-87q-14d-119cx & 87 & 14 & 119 \\
\rowcolor[gray]{0.9} l-79q-21d-112cx & 79 & 21 & 112 \\
\rowcolor[gray]{0.9} l-83q-21d-129cx & 83 & 21 & 129 \\
\rowcolor[gray]{0.9} l-86q-22d-159cx & 86 & 22 & 159 \\
\rowcolor[gray]{0.9} l-86q-29d-223cx & 86 & 29 & 223 \\
\botrule
\end{tabular}
\end{table}

\section{Performance Evaluation}
\label{sec:perf_eval}
In this section, we evaluate QARMA and QARMA-R using standard benchmarks and randomly generated quantum circuits.
We compared our QARMA and QARMA-R against a QUBO-based mapping method and IBM's Qiskit transpiler, employing its most rigorous optimization settings, measuring inter-core communication reduction and computational efficiency on a modular quantum computer architecture. 
The following subsections present a detailed analysis of our method's superior performance across varying circuit complexities and the transformative impact of qubit reuse on allocation quality.

\begin{figure*}[t]
    \centering
    \includegraphics[width=0.9\textwidth]{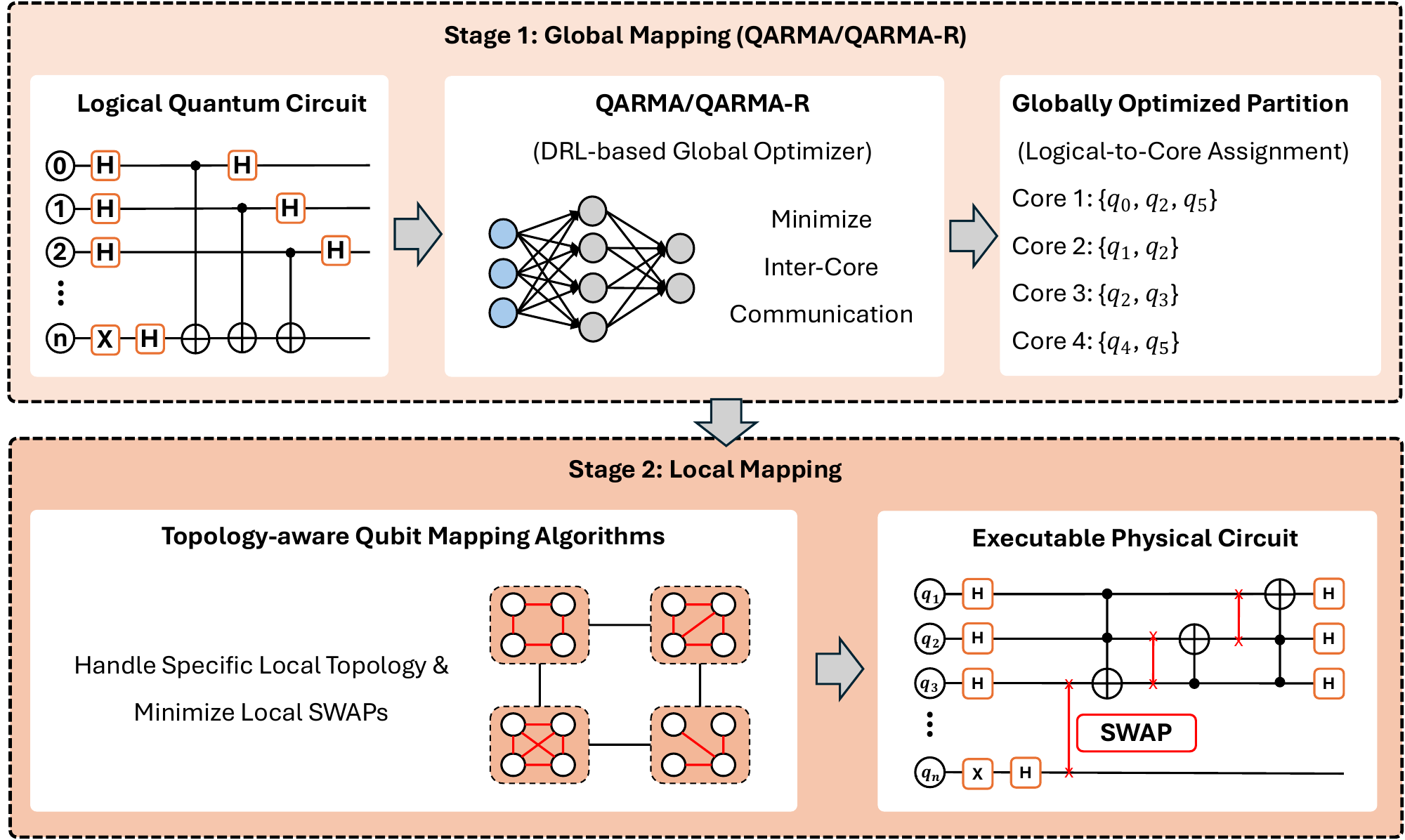}
    \caption{Hierarchical compilation pipeline. QARMA/QARMA-R (Stage 1) performs global mapping in inter-core topology to minimize costly inter-core transfers. Stage 2 uses a local qubit mapping algorithm to route within specific physical core topologies.}
    \label{fig:2stages_deployment}
\end{figure*}
\subsection{Experimental Setup}
\label{sec:exp_setup}

Our proposed QARMA and QARMA-R were evaluated on a high-performance computing system using an AMD Ryzen Threadripper PRO 5955WX 16-core CPU, 512GB of memory, and three NVIDIA RTX 4090 GPUs.

To ensure a comprehensive and rigorous evaluation, our methodology utilizes a two-part benchmark suite detailed in Table~\ref{tab:combined-circuits}. The first part consists of a diverse set of standard benchmark circuits selected from established quantum computing libraries, including many from the QASMBench suite~\cite{qasmbench}, which are common in the quantum compilation literature. These circuits (white rows in Table~\ref{tab:combined-circuits}) cover a range of algorithm types and provide a basis for comparison with prior work. The second part of our suite comprises randomly generated circuits (gray rows in Table~\ref{tab:combined-circuits}) designed to systematically assess the scalability and robustness of our framework across a broad spectrum of complexities. This dual approach allows us to validate QARMA's performance on realistic applications while also probing its behavior as circuit parameters like qubit count, depth, and gate density are scaled.

For our evaluation, we defined a series of modular quantum architectures to analyze the scalability of our approach across different module sizes. All architectures consist of 10 chips organized in a 2 $\times$ 5 grid topology. To study the impact of core capacity, we created five distinct configurations by varying the number of physical qubits per chip: 10, 20, 40, 60, and 90, each with all-to-all intra-chip connectivity. Our baseline module size of 10 qubits was deliberately chosen because smaller quantum chips are known to have significantly higher manufacturing yields due to a lower probability of fabrication defects. For example, recent architectural studies modeling state-of-the-art fabrication parameters show that 10-qubit chiplets can achieve a collision-free yield of 85.27\%, making them a practical and high-yield building block for larger, scalable quantum systems~\cite{smith2022scaling}. This set of configurations provides a representative yet challenging testbed for evaluating compilation strategies. We specifically chose a grid topology because its constrained connectivity creates a non-trivial optimization landscape where inter-core communication costs are highly dependent on the physical placement of logical qubits. Although we test on a specific grid topology, the QARMA framework is fundamentally topology-agnostic. The DRL agent learns an optimal policy subject to the provided architectural constraints, thereby remaining flexible enough to be applied to a wide variety of modular systems without altering the core methodology.

IBM's Qiskit is primarily designed for monolithic processors; however, its transpiler is highly flexible and noise-aware, making decisions based on a detailed hardware cost model. To establish a robust baseline, we customized the environment to mirror our modular architecture. We represented the entire 10-chip system as a single, large device with a unified coupling graph. Within this graph, we assigned the properties of the connections to reflect the physical reality of a modular system:
\begin{itemize}
    \item \textbf{Intra-chip links} (connections between qubits on the same chip) were configured with low error rates and short gate execution times, representing fast, high-fidelity local operations.
    \item \textbf{Inter-chip links} (connections between qubits on different chips) were configured with significantly higher error rates and longer execution times, modeling the high cost and latency of quantum state transfers between modules.
\end{itemize}
This configuration is crucial because Qiskit's aggressive optimization (\texttt{optimization\_level=3}) uses a suite of sophisticated passes, such as the \texttt{SABRE} routing algorithm, which strive to map the circuit while minimizing a cost function that is heavily influenced by gate errors and durations. By making inter-chip operations deliberately ``expensive,'' we create a strong incentive for the transpiler to avoid them whenever possible. In its effort to find the mapping with the highest overall fidelity, the algorithm naturally learns to confine operations within individual chips, thus implicitly minimizing inter-core communication. This approach effectively guides Qiskit's powerful heuristics to tackle the central challenge of modular systems, establishing a highly competitive and realistic baseline for our comparison.

The proposed method was evaluated in two configurations: reuse-disabled (\textit{QARMA}) and reuse-enabled (\textit{QARMA-R}).
The performance was benchmarked against Qiskit (v1.4.0) using its highest preset transpilation optimization level (`optimization\_level = 3', or `-O3`) and a representative QUBO-based mapping approach \cite{bandic2023mapping}. It is crucial to note that Qiskit-O3 is its most aggressive compilation strategy, applying a complex sequence of dozens of optimization passes.
This goes far beyond simple qubit mapping or routing, and includes extensive circuit synthesis techniques (e.g., gate cancellation, basis change optimization, gate fusion), advanced layout selection heuristics, sophisticated routing algorithms, and further post-layout optimizations.
Therefore, comparing against Qiskit-O3 establishes a high standard, evaluating our DRL approach against a comprehensive state-of-the-art compilation pipeline.

\begin{table}[t]
\caption{Comparison of inter-operation qubit communication counts among QUBO mapping, Qiskit 1.4.0 with optimization level 3 (-O3), and our methods in two settings: \textit{QARMA}, with qubit reuse disabled, and \textit{QARMA-R}, with reuse enabled, for selected circuits from Table~\ref{tab:combined-circuits}.}
\label{tab:combined-comparison}
\centering
\begin{tabular}{@{}l c c c c@{}}
\toprule
\toprule
\textbf{Circuit Name} & \textbf{QUBO} & \textbf{Qiskit}  & \textbf{QARMA} & \textbf{QARMA-R} \\
\toprule
4mod5-bdd-287       & 1,183 & 1 & 0  & 0  \\
ex1-226             & 106 & 2 & 0   & 0   \\
graycode6-47        & 100 & 2 & 0   & 0   \\
xor5-254            & 112 & 2 & 0   & 0   \\
bv-n10              & 203 & 3 & 2  & 0   \\
cc-n10              & 155 & 3 & 2  & 0   \\
cnt3-5-179          & 2,931 & 13 & 19  & 0  \\
multiply-n13        & 280 & 31 & 15  & 0  \\
sym9-146            & 5,440 & 122 & 69 & 0 \\
rd53-311            & 4,830 & 19 & 29 & 0 \\
cnt3-5-180          & 8,100 & 22 & 62 & 0 \\
system9             & 5,469 & 92 & 69 & 0 \\
wim-266             & 16,592 & 57 & 44 & 0 \\
sym6-316            & 4,645 & 75 & 45 & 0 \\
rd84-142            & 5,653 & 41 & 29 & 0 \\
multiplier-n15      & 1,448 & 50 & 24  & 11  \\
ising-model-13      & 11,382 & 105 & 70  & 20 \\
ising-model-16      & 13,991 & 97 & 70  & 20 \\
cm152a-212          & - & 99 & 46 & 0 \\
square-root-n18     & 8,366 & 235 & 78 & 66 \\
multiplier-n45     & - & 1,202 & 452 & 293 \\
\midrule
\rowcolor[gray]{0.9} s-10q-16d-13cx & 270 & 1 & 0 & 0 \\
\rowcolor[gray]{0.9} s-30q-5d-9cx   & 488 & 4 & 3 & 0 \\
\rowcolor[gray]{0.9} s-18q-10d-8cx  & 283 & 6 & 4 & 0 \\
\rowcolor[gray]{0.9} s-26q-9d-15cx  & 625 & 7 & 5 & 0 \\
\rowcolor[gray]{0.9} s-26q-9d-14cx  & 534 & 8 & 6 & 0 \\
\rowcolor[gray]{0.9} m-48q-7d-17cx  & 881 & 9 & 5 & 0 \\
\rowcolor[gray]{0.9} m-45q-8d-29cx  & 1,010 & 8 & 5 & 0 \\
\rowcolor[gray]{0.9} m-59q-11d-38cx & 1,663 & 33 & 19 & 13 \\
\rowcolor[gray]{0.9} m-53q-18d-48cx & 2,786 & 38 & 28 & 24 \\
\rowcolor[gray]{0.9} m-60q-15d-67cx & 3,780 & 74 & 63 & 51 \\
\rowcolor[gray]{0.9} l-87q-14d-119cx & 6,607 & 184 & 108 & 97 \\
\rowcolor[gray]{0.9} l-79q-21d-112cx & 4,454 & 182 & 69 & 56 \\
\rowcolor[gray]{0.9} l-83q-21d-129cx & 5,845 & 246 & 145 & 128 \\
\rowcolor[gray]{0.9} l-86q-22d-159cx & 7,949 & 298 & 163 & 154 \\
\rowcolor[gray]{0.9} l-86q-29d-223cx & 10,233 & 464 & 273 & 256 \\
\botrule
\end{tabular}
\end{table}

\subsection{Hierarchical Compilation Strategy}
\label{subsec:hierarchical_strategy}

It is important to clarify that QARMA and QARMA-R are designed as high-level mappers within a hierarchical compilation pipeline. As shown in Fig.~\ref{fig:2stages_deployment}, we assume an abstract all-to-all topology within each core. This design choice allows us to mathematically isolate the problem of \textit{inter-core} communication, which is the dominant source of error in modular architectures. Attempting to solve both global inter-core and local intra-core constraints simultaneously within the RL agent would drastically increase the state space and obscure the specific benefits of our inter-core optimization. By assuming all-to-all connectivity at the core level, we effectively isolate and maximize the performance of the inter-core mapping strategy.

In a real-world deployment, our framework operates as \textbf{Stage 1}: allocating logical qubits to physical cores to minimize global state transfers (inter-core operations). \textbf{Stage 2} would subsequently employ a Topology-aware Qubit Mapping Algorithm, such as SABRE~\cite{li2019tackling}, HA~\cite{HA}, or other heuristic approaches~\cite{zhang2021time, siraichi2018qubit, zulehner2018efficient}, to handle the specific local topology (e.g., Heavy-Hex or Grid) and minimize local SWAP operations within each core.

We explicitly note that we do not conduct experiments comparing intra-core SWAP operation counts in this paper. This is because the problem of single-core qubit mapping and SWAP minimization has been extensively studied and optimized in prior works~\cite{li2019tackling, HA, zhang2021time, siraichi2018qubit, zulehner2018efficient, nassc}. Our contribution focuses on the challenge of minimizing inter-core state transfers, which incur errors orders of magnitude higher than local SWAPs. By decoupling these optimization problems, QARMA and QARMA-R provide a globally optimized partition that these standard mapping algorithms can then route locally.

\subsection{Performance Results}
\label{sec:results}

We evaluated the primary optimization objective, minimizing inter-operation qubit communication counts, by comparing the number of required inter-core qubit state transfers for different mapping strategies. We benchmarked our proposed methods, \textit{QARMA} (without reuse) and \textit{QARMA-R} (with reuse), against a representative QUBO-based approach~\cite{bandic2023mapping} and IBM's Qiskit transpiler at its highest optimization level (O3).

A comprehensive breakdown of the inter-core connection counts for all benchmark circuits across the full range of tested architectures is provided in Appendix~\ref{appendix:detail_inter-core_benchmark}. The following sections summarize these results and highlight the significant performance gains achieved by our methods.

\begin{figure*}[t] 
    \centering 
    
    \begin{subfigure}[b]{0.48\textwidth}
        \centering
        \includegraphics[width=\textwidth]{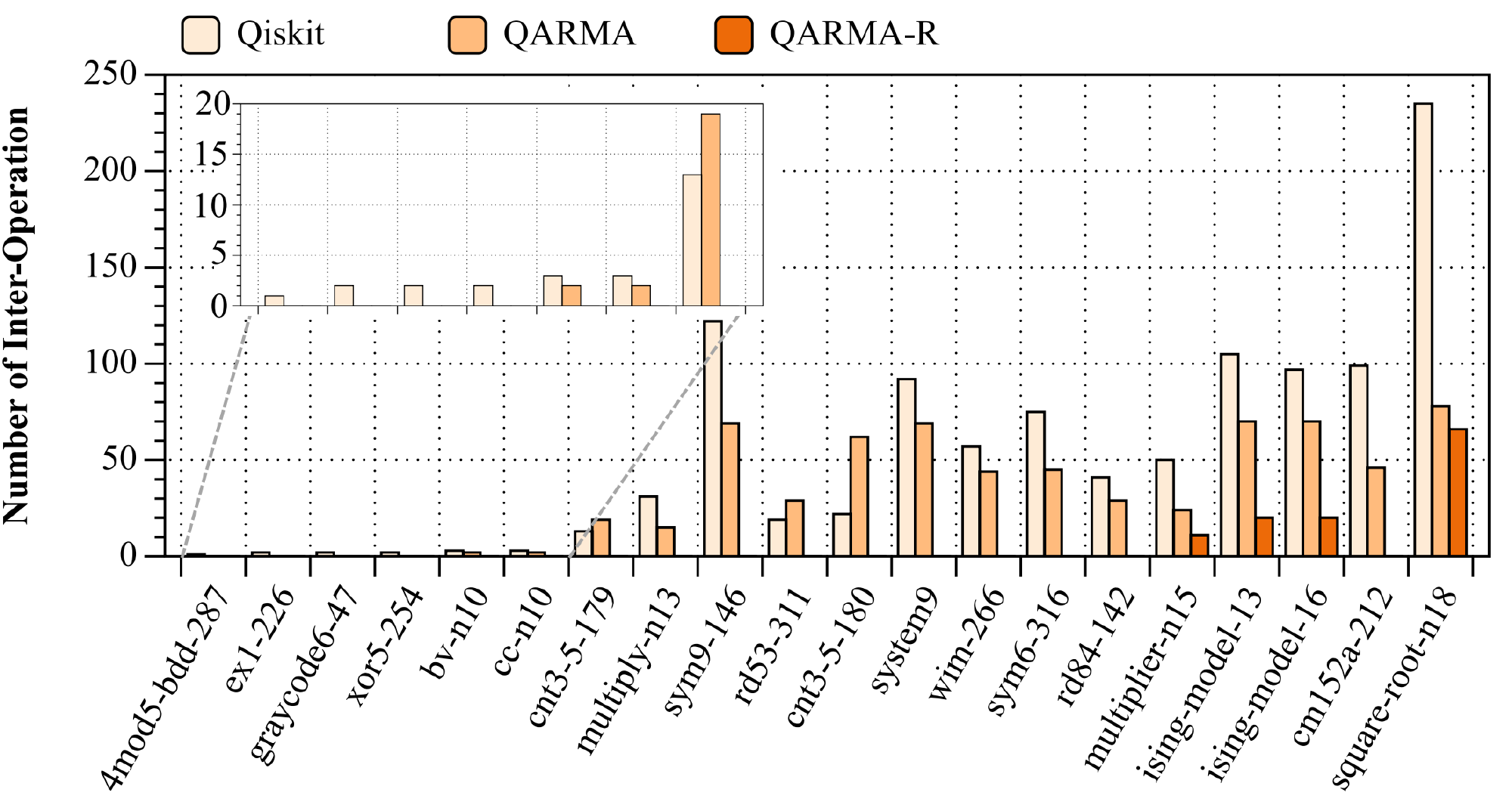}
        \caption{Standard circuits from Table~\ref{tab:combined-circuits} white rows.}
        \label{fig:first}
    \end{subfigure}
    \begin{subfigure}[b]{0.49\textwidth}
        \centering
        \includegraphics[width=\textwidth]{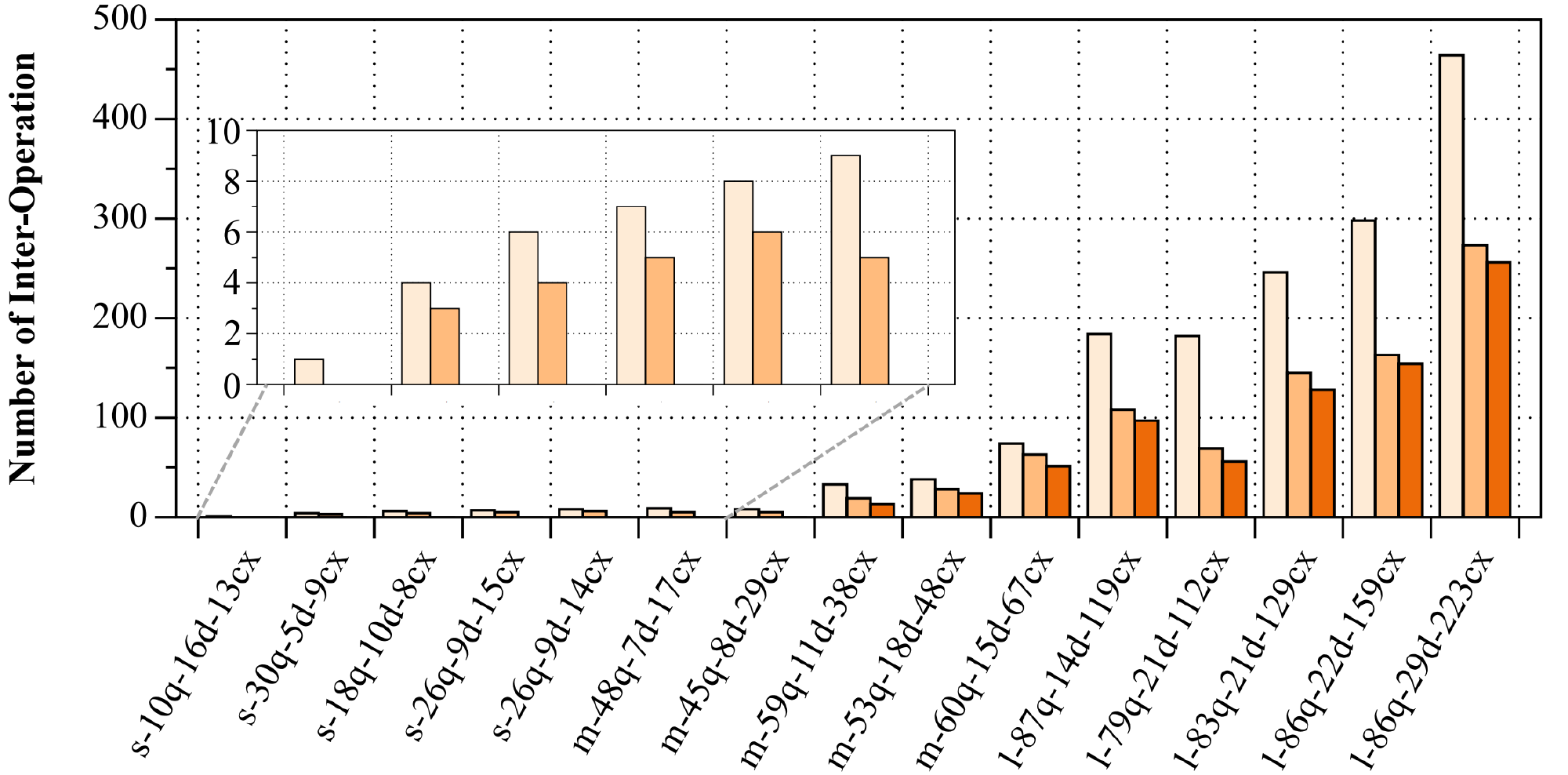}
        \caption{Random circuits from Table~\ref{tab:combined-circuits} gray rows.}
        \label{fig:second}
    \end{subfigure}
    \caption{Comparison of inter-operation qubit communication or inter-core connection counts using Qiskit. The baseline uses Qiskit 1.4.0, with the highest optimization level (Level 3). Bars show the absolute number of inter-operation counts. Our method was evaluated in two settings: \textit{QARMA}, with qubit reuse disabled, and \textit{QARMA-R}, with reuse enabled. The inset highlights results for smaller circuits.}
\label{fig:benchmark_result}
\end{figure*}

\subsubsection{Comparison with QUBO-based Mapping}
\label{sec:qubo-comparison}

Table~\ref{tab:combined-comparison} highlights the substantial advantage of our approach over the QUBO-based mapping approach described in Section~\ref{sec:related_work}.
For nearly all benchmark circuits for which QUBO results are available, our QARMA and QARMA-R achieve significantly fewer inter-core connections. 
The \textit{QARMA-R} configuration demonstrates the most dramatic improvements, eliminating inter-core communications (0 connections) for a large majority of the standard benchmarks, e.g., \textit{4mod5-bdd-287}, \textit{ex1-226}, \textit{bv-n10}, \textit{cc-n10}, \textit{cnt3-5-179}, \textit{multiply-n13}, \textit{sym9-146}, \textit{rd53-311}, \textit{cnt3-5-180}, \textit{system9}, \textit{wim-266}, \textit{sym6-316}, \textit{rd84-142}, and \textit{cm152a-212}. 
This contrasts starkly with the QUBO method, which often requires hundreds or even thousands of connections for the same circuits, for example, 1,183 for \textit{4mod5-bdd-287}, 5,440 for \textit{sym9-146}, and 16,592 for \textit{wim-266}. 
Quantitatively, the reduction achieved by our \textit{QARMA-R} method compared to QUBO typically ranges from 98\% to 100\% across both standard and random benchmarks.

Even without exploiting reuse (\textit{QARMA}), our core allocation strategy consistently outperformed QUBO, although with a smaller margin than that of the reuse-enabled version. For instance, on \textit{sym9-146}, the \textit{QARMA} method uses only 69 connections compared with QUBO's 5,440 (>99\% reduction). 
Similar significant reductions were observed for other complex circuits such as \textit{cnt3-5-180} (19 vs. 8,100) and \textit{wim-266} (44 vs. 16,592).

The performance gap remains substantial for the randomly generated circuits. 
For the largest circuit, i.e., \textit{l-86q-29d-223cx}, the QUBO method requires 10,233 connections, whereas our \textit{QARMA-R} method requires only 256 (97.50\% reduction), and the \textit{QARMA} method requires 273 (97.33\% reduction). 
This significant difference arises from the QUBO method's fundamentally local, slice-by-slice optimization. While it effectively groups ``friend qubits'' for gates within a single time-slice, it lacks a global perspective on the circuit's entire dependency structure. Consequently, a mapping that is optimal for one slice may be highly suboptimal for future slices, forcing numerous costly inter-core operations later in the execution. In contrast, our DRL-based approach learns the long-range dependencies across the entire circuit, allowing it to make globally aware placement decisions that anticipate future interactions and minimize the total communication overhead.
These results suggest that our approach is better equipped to handle the complex structural dependencies and optimization landscape inherent in qubit mapping, particularly for larger circuits. 
Furthermore, the QUBO formulation, as described in \cite{bandic2023mapping}, does not inherently incorporate qubit reuse, preventing it from leveraging this powerful resource optimization technique, which our approach effectively integrates.

\subsubsection{Comparison with Qiskit Mapping}
\label{sec:qiskit-comparison}

Appendix~\ref{appendix:detail_inter-core_benchmark} presents a detailed breakdown of inter-core connection counts for all benchmark circuits across the entire range of tested architectures. For clarity, we summarize the results in Fig.~\ref{fig:benchmark_result} and Table~\ref{tab:combined-comparison} to facilitate comparison with Qiskit, customized for modular architecture with highly optimized, i.e., -O3, transpilation pipeline, which encompasses numerous circuit optimization and mapping techniques beyond basic routing.
Even against this strong baseline, our approach demonstrates significant advantages, particularly when qubit reuse is enabled (\textit{QARMA-R}).

The \textit{QARMA-R} configuration consistently requires fewer inter-core connections compared with Qiskit-O3 across almost all benchmarks. For many standard circuits, such as \textit{4mod5-bdd-287}, \textit{ex1-226}, \textit{bv-n10}, \textit{multiply-n13}, and \textit{sym9-146}, our reuse-enabled method \textit{QARMA-R} reduces the inter-core connection count to zero.
In contrast, Qiskit still requires a small number (1-122 connections). For more complex circuits, where Qiskit requires a substantial number of connections, our \textit{QARMA-R} method offers notable reductions.
For example, on \textit{multiplier-n15}, connections are reduced from 50 (Qiskit) to 11 (\textit{QARMA-R}); on \textit{ising-model-13}, reductions are from 105 to 20, respectively.
The benefit is also evident on \textit{square-root-n18} (235 down to 66) and \textit{multiplier-n45} (1,202 down to 293).

Even without the reuse mechanism (\textit{QARMA}), the DRL method demonstrates performance comparable to the highly optimized Qiskit baseline.
For simple circuits such as \textit{ex1-226} and \textit{bv-n10}, the results are similar to Qiskit's. Furthermore, the \textit{QARMA} DRL approach shows a distinct advantage on more complex circuits, particularly the larger randomly generated benchmarks.
For instance, on circuits like \textit{multiply-n13}, \textit{wim-266}, \textit{rd84-142}, and the large `l-' scale random benchmarks, the \textit{QARMA} method without reuse achieves fewer inter-core connections than the Qiskit-O3 pipeline.
This suggests that the learned policy, even without explicitly optimizing for reuse, can find efficient mappings that outperform complex heuristic strategies on larger, more intricate problem instances.

It is important, however, to analyze the instances in Fig.~\ref{fig:benchmark_result}a where \textit{QARMA} fails to outperform the baseline, resulting in a higher number of inter-core connections (e.g., for circuits like \textit{cnt3-5-179} and \textit{rd53-311}. We hypothesize that this occurs because the global optimization strategy learned by the DRL agent may not offer an advantage on smaller or structurally simpler circuits. In these cases, Qiskit’s highly tuned, aggressive heuristics can be more effective at finding a locally optimal mapping to reduce interoperation costs. The complexity and overhead of the DRL model are well-suited to navigating the vast, intricate search spaces of larger circuits, where its ability to learn global patterns provides a distinct advantage over more localized heuristic approaches.

Overall, the results across standard and randomly generated benchmarks confirm the effectiveness of our QARMA and QARMA-R.
The reuse-enabled (\textit{QARMA-R}) strategy consistently provides the best results, achieving substantial reductions compared to Qiskit-O3 across diverse circuit types based on Table~\ref{tab:combined-circuits}, averaging an 86\% reduction in inter-core communication and reaching up to 100\% reduction (i.e., zero connections) in many cases.
Even the reuse-disabled (\textit{QARMA}) approach demonstrates notable benefits for larger circuits, yielding improvements between 15-40\% over Qiskit-O3 for these instances.
This highlights the combined power of the DRL's learned allocation strategy and the integration of the qubit reuse mechanism in minimizing costly inter-core communication, significantly outperforming even highly optimized classical compilation pipelines, especially as circuit size and complexity grow.

\subsubsection{Computational Efficiency}
\label{sec:computational-efficiency}

While our primary objective is to minimize inter-core communication, execution time is another critical factor in practical quantum circuit mapping.
We comprehensively analyzed computational efficiency across all methods, as execution time directly impacts compiler usability in quantum computing workflows.

Table~\ref{tab:execution-time-comparison} summarizes the execution time ranges for the four mapping approaches across our benchmark suite. The results reveal significant performance differences between the methods:

\begin{table}[t]
\caption{Execution time ranges for circuit mapping across different approaches. Small circuits ($<$10 two-qubit gates), Medium circuits (10-50 gates), Large circuits ($>$50 gates).}
\label{tab:execution-time-comparison}
\centering
\begin{tabular}{@{}l c c c@{}}
\toprule
\toprule
\textbf{Method} & \textbf{Small} & \textbf{Medium} & \textbf{Large} \\
\toprule
QUBO & 6.7s--27.1s & 13.3s--1,887s & 275s--54,567s \\
\midrule
Qiskit-O3 & 0.15s--0.82s & 0.83s--3.10s & 1.1s--10.7s \\
\midrule
QARMA & 0.15s--2.12s & 0.28s--6.14s & 12.1s--614.9s \\
\midrule
QARMA-R & 0.34s--2.30s & 0.75s--9.31s & 13.6s--644.1s \\
\bottomrule
\bottomrule
\end{tabular}
\end{table}

Our analysis reveals several key findings:

\begin{itemize}
    \item \textbf{QUBO scaling limitations:} The QUBO approach, implemented using the simulated annealing solver from qubovert \cite{tiosue2020qubovert}, exhibits severe computational scaling issues. Execution times increase dramatically for complex circuits, with mapping \textit{wim\_266.qasm} requiring over 15 hours, compared to just 115.3 seconds with our reuse-enabled approach, i.e., 470$\times$ speedup.
    
    \item \textbf{DRL efficiency:} Despite employing sophisticated neural network inference, our QARMA and QARMA-R methods maintain reasonable execution times across the benchmark suite. For most circuits, our approach executes in seconds or minutes, making it practical for real-world quantum compilation workflows.
    
    \item \textbf{Reuse overhead:} The additional computation required for qubit reuse analysis introduces modest overhead in execution time (typically 10-30\% longer than non-reuse). However, this increase is insignificant compared to the dramatic reduction in inter-core communications.
    
    \item \textbf{Scaling with circuit complexity:} All methods show increased execution times for larger circuits, but with vastly different scaling characteristics. Our measurements for the QUBO solver, running the \texttt{anneal\_qubo} function, exhibit near-exponential scaling with circuit size and gate count, while our DRL methods maintain more reasonable polynomial scaling relationships.
\end{itemize}

While Qiskit consistently offers the fastest compilation times, our approach presents a strategic trade-off between compilation speed and the quality of the final mapped circuit. Although our methods can be 10-60x slower than Qiskit for large circuits, we argue this higher, one-time classical compilation cost is a worthwhile investment. The resulting mappings dramatically reduce inter-core communication—often the most significant source of errors and latency in modular quantum systems. On expensive and noisy quantum hardware, minimizing these costly operations is paramount for achieving high-fidelity results. Therefore, investing additional time in compilation to produce a superior circuit mapping can lead to a much higher probability of successful execution, making our approach highly practical for real-world NISQ-era workflows.

It is worth noting that the QUBO performance could potentially be improved with specialized hardware solvers or more efficient classical implementations.
However, our experiments using the standard simulated annealing approach in \texttt{qubovert.sim} demonstrate the inherent complexity of the quadratic formulation that makes scaling difficult, even with optimized solvers.

The execution time analysis confirms that our QARMA and QARMA-R approaches balance solution quality and computational efficiency.
Our method delivers near-optimal allocations within reasonable timeframes for circuits of practical interest, making it suitable for integration into quantum computing workflows.
This efficiency, combined with the dramatic reduction in inter-core communication shown in Sections~\ref{sec:qubo-comparison} and~\ref{sec:qiskit-comparison}, establishes our approach as a significant advancement in quantum circuit mapping for modular architectures.

\begin{figure*}[t] 
    \centering 
    
    \begin{subfigure}[b]{0.8\textwidth}
        \centering
        \includegraphics[width=\textwidth]{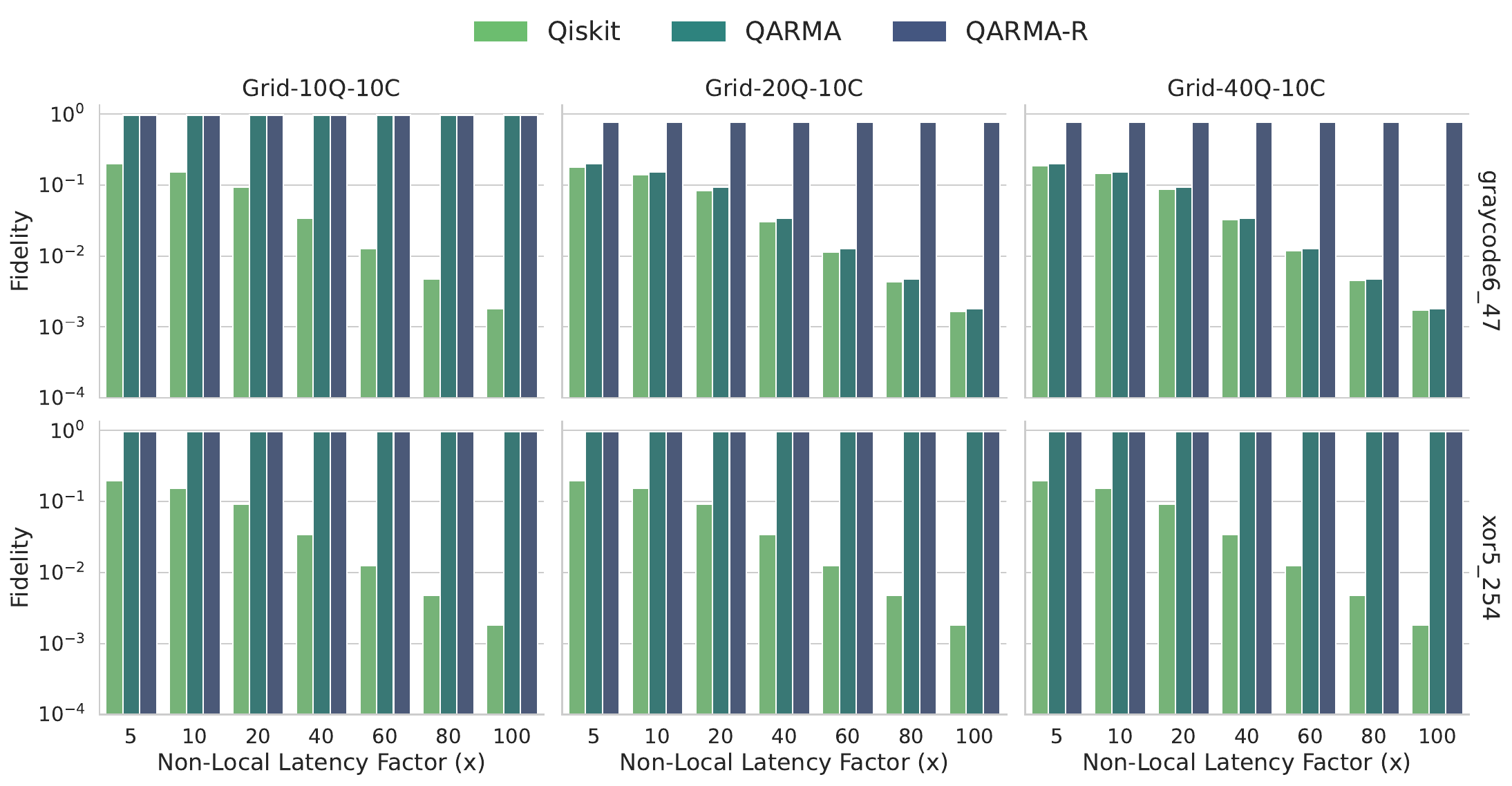}
        \caption{Standard circuits from Table~\ref{tab:combined-circuits} white rows.}
        \label{fig:first}
    \end{subfigure}
    \begin{subfigure}[b]{0.8\textwidth}
        \centering
        \includegraphics[width=\textwidth]{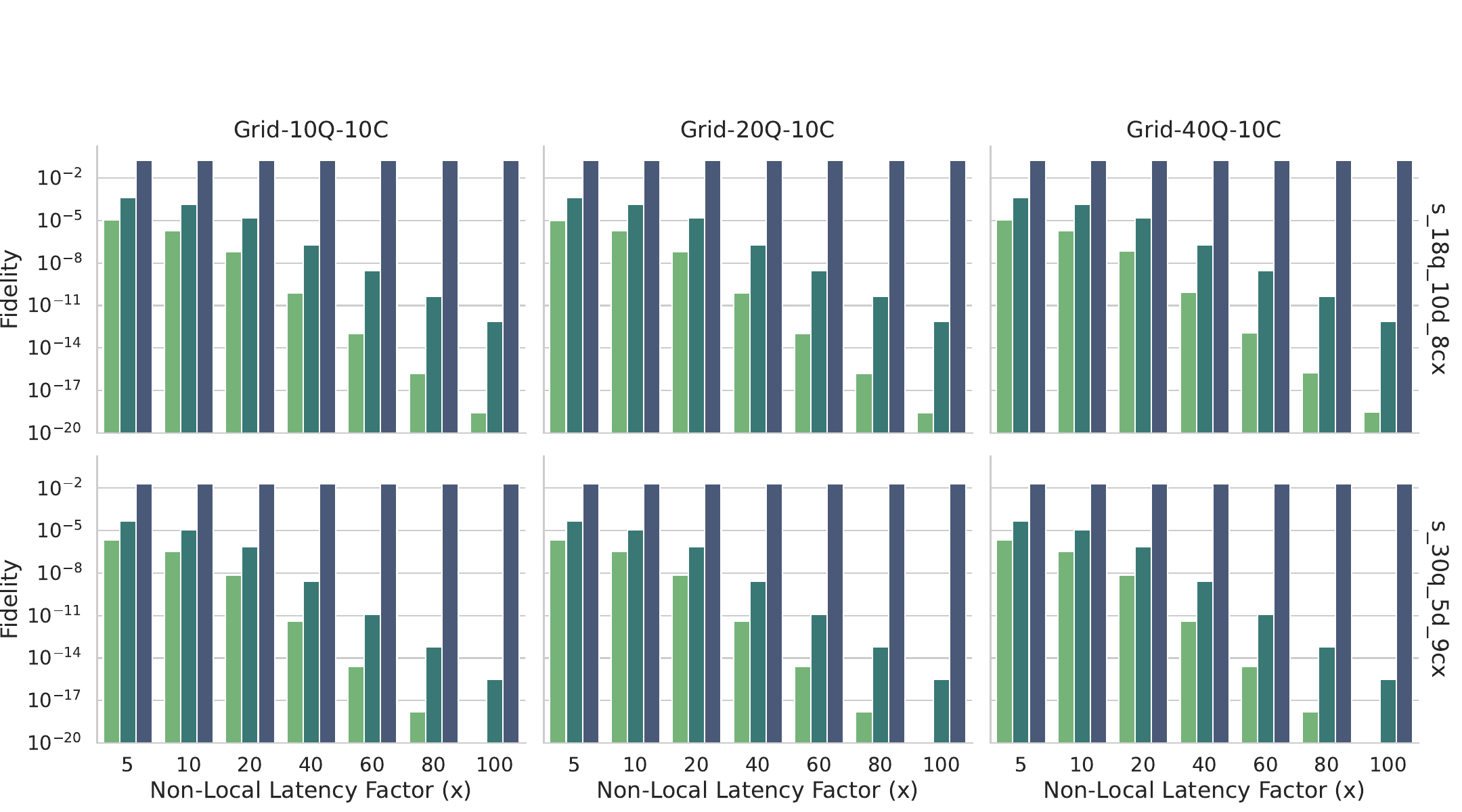}
        \caption{Random circuits from Table~\ref{tab:combined-circuits} gray rows.}
        \label{fig:second}
    \end{subfigure}
    
    \caption{Fidelity comparison of Qiskit, QARMA, and QARMA-R across different modular architectures and non-local communication latency factors.}
    \label{fig:fidelity_standard_and_random}
\end{figure*}

\subsection{Fidelity Impact of Qubit Reuse vs. Inter-Core Connections}
A key feature of QARMA-R is its integration of a dynamic qubit reuse mechanism that strategically inserts mid-circuit measurement and reset operations. While this technique is effective at reducing the total number of physical qubits required, it can increase the overall circuit depth. This raises a critical question: Does the fidelity penalty from increased depth negate the benefits of reuse? Our analysis indicates that in the context of modular architectures, the answer is a firm no. The fidelity cost of inter-core communication is substantially more severe than the cost of the additional operations required for reuse.

To quantify this trade-off, we conducted a comparative fidelity analysis between QARMA (without reuse), QARMA-R (with reuse), and Qiskit. We employed a robust fidelity estimation model adapted for modular systems, based on the analytical framework proposed by Escofet et al. \cite{fidelity_estimation}. This model treats each inter-core state transfer not merely as a SWAP, but as a high-latency operation with a specific error rate. For instance, the transfer of a qubit via a quantum matter-link can take up to $400\,\mu\text{s}$ per hop \cite{titan-framework}. Experimental error budgets for quantum gate teleportation, a common method for inter-module operations, show a predicted total error of $\mathbf{12.1\%}$ \cite{distribute_quantum_optical_network}. The dominant error sources in this process are the initial remote entanglement ($3.11\%$) and the local mixed-species gates ($2.0\%-2.4\%$), which are orders of magnitude higher than the error from the mid-circuit measurements ($0.091\%-0.122\%$) that enable qubit reuse.

Our fidelity model captures these physical realities through a series of analytical steps. First, it translates the physical gate fidelity ($F_g$) from device calibration data into a theoretical depolarization parameter ($p$) for a $d$-dimensional quantum system (where $d=2^n$ for $n$ qubits), using the following relation:
\begin{equation}
    p = \frac{d(1 - F_g)}{d - 1}
    \label{eq:depolarization}
\end{equation}
This allows the model to uniformly represent different sources of gate error within the depolarizing channel framework.

The fidelity of each individual qubit ($F_q$) is then updated iteratively as gates are applied. For a single-qubit gate acting on qubit $q_i$, the new fidelity $F'_{q_i}$ is an affine transformation of its previous fidelity $F_{q_i}$:
\begin{equation}
    F'_{q_i} = (1-p)F_{q_i} + (1-p_{ent})\frac{p}{d}
    \label{eq:single_qubit_update}
\end{equation}
Here, the term $(1-p)F_{q_i}$ represents the decay of the initial fidelity, while the second term captures the admixture of a fully mixed state, scaled by the gate's error $p$. The hyperparameter $p_{ent} \in [0, 1]$ accounts for the system's entanglement, as a more entangled state is more susceptible to fidelity loss from local depolarization.

For a two-qubit gate acting on qubits $q_i$ and $q_j$, the fidelity update is more complex, as the error is correlated. An intermediate error distribution term, $\eta$, is first calculated:
\begin{equation}
    \begin{split}
        \eta = \frac{1}{2}\Bigg( & -\sqrt{1-p}(F_{q_i} + F_{q_j}) + {} \\\\
        & \sqrt{(1-p)(F_{q_i} + F_{q_j})^2 + \frac{4p}{d_{AB}} \vphantom{F_{q_i}}} \Bigg)
    \end{split}
    \label{eq:eta_term}
\end{equation}

This term quantifies the extent to which the depolarization error is shared between the two interacting qubits, depending on their prior fidelities. The individual fidelities are then updated as:
\begin{align}
    F'_{q_i} &= \sqrt{1-p} \cdot F_{q_i} + (1-p_{ent}) \cdot \eta \label{eq:two_qubit_update1} \\
    F'_{q_j} &= \sqrt{1-p} \cdot F_{q_j} + (1-p_{ent}) \cdot \eta \label{eq:two_qubit_update2}
\end{align}
These equations show that each qubit's new fidelity is a combination of its scaled previous fidelity and this shared error term $\eta$.

Crucially, during the long duration ($t_{op}$) of a state transfer or a slow gate, all other qubits in the system remain idle. This idle time is a primary source of fidelity loss in modular systems. We model this system-wide decoherence by applying an exponential decay factor to every qubit's fidelity after each time slice, based on its intrinsic coherence times ($T_1$ for relaxation and $T_2$ for dephasing):
\begin{equation}
    F'_{q} = F_{q} \cdot e^{-t_{op}/T_1^q} \cdot \left(\frac{1}{2}e^{-t_{op}/T_2^q} + 0.5\right)
    \label{eq:decoherence}
\end{equation}

To substantiate this claim, we conducted a comprehensive fidelity analysis comparing QARMA-R, QARMA, and the highly optimized Qiskit baseline, with detailed results presented in Appendix~\ref{appendix:detail_fidelity_benchmark}. Our fidelity estimation model was configured with parameters mimicking a state-of-the-art superconducting device, specifically using the best-case values from IBM's \texttt{ibm\_brisbane} machine: a T1 of 335$\mu$s, T2 of 205$\mu$s, a single-qubit gate error of 1e-4 with a 10ns duration, and a two-qubit (ECR) gate error of 2.75e-3 with a 590ns duration.

Crucially, our model accounts for the fact that non-local communication operations are not only slower but also significantly more error-prone. As a real-world reference, the fidelity of a teleported CZ gate between two modules separated by about two meters was measured to be 86.2\%. The fidelity of the raw remote entanglement between the network qubits was 96.89\%~\cite{distribute_quantum_optical_network}. Informed by this, we conservatively set the fidelity cost for a single inter-core state transfer in our model to 96\%. The analysis, summarized in Appendix~\ref{appendix:detail_fidelity_benchmark},  Fig.~\ref{fig:fidelity_standard_and_random}a and Fig.~\ref{fig:fidelity_standard_and_random}b, evaluates performance across four modular architectures (scaling from 10 to 90 qubits per core) and under varying non-local communication costs, represented by a multiplicative latency factor from 5x to 100x that of a local gate~\cite{baker2020time}.

The results unequivocally demonstrate the superiority of the reuse-enabled strategy. As shown in Fig.~\ref{fig:fidelity_standard_and_random}, for the standard benchmarks \texttt{graycode6\_47} and \texttt{xor5\_254}, QARMA-R maintains a near-perfect execution fidelity across all tested architectures and latency factors. This is a direct result of its ability to eliminate inter-core state transfers entirely. In stark contrast, the fidelity of both QARMA and Qiskit, which incur one or more state transfers, collapses precipitously as the communication latency increases. For these methods, a latency factor of 40x or higher renders the circuits effectively unrunnable, with fidelity dropping by orders of magnitude.

This trend is even more pronounced for the more complex, randomly generated circuits shown in Fig.~\ref{fig:fidelity_standard_and_random}b. While the absolute fidelity is lower for all methods due to the circuits' intrinsic complexity, QARMA-R consistently achieves a fidelity that is several orders of magnitude higher than the alternatives. For the \texttt{s\_30q\_5d\_9cx} circuit, QARMA-R's fidelity is relatively stable, whereas the fidelity of QARMA and Qiskit quickly plummets to near-zero ($<10^{-15}$). This highlights that as inter-core communication becomes more expensive—a realistic scenario in near-term hardware—the ability to avoid it through qubit reuse becomes the single most critical factor for successful execution.

This analysis validates the core design principle of QARMA-R: in a modular quantum architecture, the primary optimization objective must be to minimize inter-core communication. The localized fidelity cost of adding measurement and reset gates is a small and acceptable price to pay for avoiding the catastrophic, system-wide decoherence induced by the high latency and error rates of quantum state transfers. Therefore, exploiting qubit reuse is not just a method for resource reduction but a critical strategy for enhancing execution fidelity on near-term modular quantum computers.

\section{Discussion}
\label{sec:discuss}

Our results demonstrate that \textit{QARMA} and its reuse-enabled extension, \textit{QARMA-R}, are highly effective at minimizing inter-core communication in modular quantum architectures. The core innovations—a transformer-based circuit encoder and a dynamic qubit reuse mechanism—were validated not only by reducing the number of costly state transfers but also through a comprehensive, hardware-calibrated fidelity analysis. This analysis confirms a critical takeaway for the field: optimizing for inter-core communication is paramount for achieving high-fidelity execution on near-term modular systems.

The fidelity estimation, grounded in parameters from IBM's \texttt{ibm\_brisbane} device, revealed a crucial trade-off. While qubit reuse can modestly increase circuit depth, as noted by Hua et al.~\cite{hua2022exploiting}, our findings show that the resulting fidelity cost is negligible compared to the catastrophic fidelity collapse caused by even a single inter-core state transfer, especially under realistic communication latencies. \textit{QARMA-R}'s ability to eliminate these transfers allows it to maintain a high and stable fidelity where other methods fail, validating that the aggressive pursuit of qubit reuse is a winning strategy.

To contextualize the performance of QARMA and QARMA-R relative to recent state-of-the-art attention-based DRL approaches, we conducted an indirect comparison with the method proposed by Russo et al.~\cite{russo2024attention}. It is important to note that their framework adopts the same architectural abstraction used in our study: they explicitly assume that physical qubits within each core are 10 all-to-all connected physical qubits, driven by the premise that inter-core communication is significantly more costly than intra-core operations. While a direct experimental comparison was not feasible due to the unavailability of their source code and the training datasets differ significantly, this shared assumption enables a meaningful analysis of their reported results on comparable Grid topologies with 10-qubit cores to highlight the advantages of our optimization strategy.
The fundamental advantage of QARMA-R is the integration of qubit reuse, which allows logical circuits to be executed on fewer physical resources. Russo et al.~\cite{russo2024attention} employ a static mapping strategy that forces the distribution of logical qubits across multiple cores when the circuit width exceeds the capacity of a single core. This limitation is evident in their reported results for the Deutsch-Jozsa algorithm, where the 50-qubit with 49-two-gates benchmark requires 125 inter-core communications on a Grid topology because the circuit width exceeds the core capacity. In contrast, QARMA-R leverages dynamic mid-circuit measurement and reset to compress the physical footprint of the circuit. For structural similarity, our medium random benchmark (\texttt{m-60q-15d-67cx}), where 60-qubit with 67-two-gates, QARMA-R achieves 51 inter-core communications (Table~\ref{tab:combined-comparison}). By reusing auxiliary qubits, QARMA-R can serialize the execution, fitting wider logical circuits into fewer physical cores, thereby eliminating the expensive inter-core state transfers that static mappers must incur.
QARMA-R demonstrates a better reduction in communication overhead. This suggests that our approach’s incorporation of dynamic qubit reuse represents a fundamental advancement addressing both communication and resource constraints critical for practical modular quantum computing.

However, our study also highlights areas for future advancement. A key limitation is that while our analysis is noise-aware, the DRL agent's reward function during training is still primarily driven by minimizing the number of inter-core connections as a proxy for fidelity. Our fidelity study demonstrates that this is a highly effective proxy; however, the agent does not directly optimize a full, hardware-specific fidelity model in real time. Furthermore, the model currently assumes uniform costs for all inter-core connections, whereas real hardware exhibits significant variations in qubit quality and link performance.

Addressing these limitations is the central priority for our future work. The next logical step is to evolve our method by developing a multi-objective DRL agent that learns to directly navigate the complex trade-offs among inter-core communication, intra-core routing, circuit depth, and resource utilization. A key enhancement will be the integration of hardware-specific noise profiles into the state representation and reward function. This will allow the agent to make more granular, noise-aware decisions—for example, choosing a high-fidelity communication link over a noisy one, or even accepting a state transfer if the only reuse option would lead to excessive decoherence from qubit idling—thereby optimizing for the highest probability of successful execution on a specific quantum device. Furthermore, to mitigate the computational overhead of our sophisticated DRL model and make it more practical for larger, more complex circuits, we will explore the use of parallel classical computing. By parallelizing the agent’s inference and decision-making, we can significantly reduce the overall mapping time, addressing the execution time trade-off noted in our efficiency analysis in Section~\ref{sec:computational-efficiency}.

\section{Conclusion}
\label{sec:conclusion}

We presented QARMA and QARMA-R, a deep reinforcement learning approach for qubit allocation with dynamic reuse in modular quantum architectures. Our results show inter-core communications reduced by up to 100\% (on average 86\%) compared to Qiskit's highest optimization level, with 15-40\% improvement for larger circuits even without reuse. Against QUBO-based mapping, our approach achieves a 97-100\% reduction with reuse enabled.

Crucially, our hardware-calibrated fidelity analysis demonstrates that the trade-off between qubit reuse and circuit depth heavily favors reuse. The catastrophic, system-wide decoherence induced by high-latency quantum state transfers is far more detrimental to execution fidelity than the negligible cost of local measurement and reset operations. Future work will build on this insight by developing a noise-aware, multi-objective compilation framework that integrates hardware-specific error rates to balance competing objectives such as inter-core communication, circuit depth, and overall execution fidelity. As quantum hardware advances, efficient and fidelity-aware compilation techniques like ours will be essential for bridging the gap between abstract algorithms and physical implementations.






\section*{Acknowledgement}
This research was supported in part by Institute for Information \& communications Technology Planning \& Evaluation (IITP) grant (No. RS-2020-II200014, A Technology Development of Quantum OS for Fault-tolerant Logical Qubit Computing Environment) and in part by Quantum Science and Technology Flagship Project (Quantum Computing) (No. RS-2025-25464760) through the National Research Foundation of Korea(NRF), funded by the Korean government (Ministry of Science and ICT(MSIT)).


\appendix

\begin{table}[t]
\caption{Weight Configurations for Sensitivity Analysis: balancing depth cost ($w_0$), earliness ($w_1$), and gap duration ($w_2$).}
\label{tab:sensitivity_weights}
\centering
\resizebox{\columnwidth}{!}{%
\begin{tabular}{@{}l l@{}}
\toprule
\toprule
\textbf{Weights $(w_0, w_1, w_2)$} & \textbf{Focus} \\
\toprule
Baseline $(1, 1, 1)$ & Balanced approach \\
\midrule
Depth-Averse $(5, 1, 1)$ & Penalizes circuit depth increase.  \\
\midrule
Reuse-Aggressive $(1, 5, 1)$ & Prioritizes the earliest finishing qubits.  \\
\midrule
Locality-Focused $(1, 1, 5)$ & Prioritizes minimizing idle gap duration.  \\
\midrule
Greedy $(0, 1, 0)$ & Ignores depth cost entirely.  \\
\bottomrule
\bottomrule
\end{tabular}
}
\end{table}

\begin{table*}[t]
\caption{Sensitivity analysis of inter-core operation counts for low- and medium-complexity benchmark circuits, all of which maintained zero inter-core operations across configurations.}
\label{tab:sensitivity_full}
\centering
\begin{tabular}{lccccc}
\toprule
\textbf{Circuit} & \textbf{Baseline} & \textbf{Depth-Averse} & \textbf{Reuse-Aggressive} & \textbf{Locality-Focused} & \textbf{Greedy} \\
 & $(1,1,1)$ & $(5,1,1)$ & $(1,5,1)$ & $(1,1,5)$ & $(0,1,0)$ \\
\midrule
4mod5-bdd\_287 & 0 & 0 & 0 & 0 & 0 \\
bv\_n10 & 0 & 0 & 0 & 0 & 0 \\
cc\_n10 & 0 & 0 & 0 & 0 & 0 \\
cm152a\_212 & 0 & 0 & 0 & 0 & 0 \\
cnt3-5\_179 & 0 & 0 & 0 & 0 & 0 \\
cnt3-5\_180 & 0 & 0 & 0 & 0 & 0 \\
ex1\_226 & 0 & 0 & 0 & 0 & 0 \\
graycode6\_47 & 0 & 0 & 0 & 0 & 0 \\
multiply\_n13 & 0 & 0 & 0 & 0 & 0 \\
rd53\_311 & 0 & 0 & 0 & 0 & 0 \\
rd84\_142 & 0 & 0 & 0 & 0 & 0 \\
sym6\_316 & 0 & 0 & 0 & 0 & 0 \\
sym9\_146 & 0 & 0 & 0 & 0 & 0 \\
system9 & 0 & 0 & 0 & 0 & 0 \\
wim\_266 & 0 & 0 & 0 & 0 & 0 \\
xor5\_254 & 0 & 0 & 0 & 0 & 0 \\
\bottomrule
\end{tabular}
\end{table*}

\begin{table*}[t]
\centering
\caption{Detailed Comparison of Inter-Operation Counts Across Multiple Modular Architectures. This table presents the raw number of inter-core state transfers required for each benchmark circuit when compiled by Qiskit (-O3), QARMA (no reuse), and QARMA-R (with reuse). The comparison is performed across five configurations of a 10-core modular architecture (arranged in a 2x5 grid), where each configuration varies the number of physical qubits per core (10Q, 20Q, 40Q, 60Q, and 90Q), demonstrating how resource availability impacts communication overhead.}
\label{tab:icc_bench_circuits}

\begin{tabular}{@{}l|rrrrr|rrrrr|rrrrr@{}}
\toprule
\toprule
\multirow{2}{*}{\textbf{Circuit}} & \multicolumn{5}{c}{\textbf{Qiskit}} & \multicolumn{5}{c}{\textbf{QARMA}} & \multicolumn{5}{c}{\textbf{QARMA-R}} \\ 
\cmidrule(lr){2-6} \cmidrule(lr){7-11} \cmidrule(lr){12-16}
& \textbf{10Q} & \textbf{20Q} & \textbf{40Q} & \textbf{60Q} & \textbf{90Q} & \textbf{10Q} & \textbf{20Q} & \textbf{40Q} & \textbf{60Q} & \textbf{90Q} & \textbf{10Q} & \textbf{20Q} & \textbf{40Q} & \textbf{60Q} & \textbf{90Q} \\
\midrule
4mod5-bdd & 1 & 0 & 0 & 0 & 0 & 0 & 0 & 0 & 0 & 0 & 0 & 0 & 0 & 0 & 0 \\
graycode6 & 2 & 1 & 2 & 2 & 2 & 0 & 1 & 1 & 1 & 1 & 0 & 0 & 0 & 0 & 0 \\
xor5 & 2 & 1 & 2 & 2 & 2 & 0 & 0 & 0 & 1 & 1 & 0 & 0 & 0 & 0 & 0 \\
bv\_n10 & 3 & 3 & 3 & 3 & 3 & 2 & 3 & 3 & 3 & 3 & 0 & 0 & 0 & 0 & 0 \\
cc\_n10 & 3 & 3 & 3 & 3 & 3 & 2 & 3 & 3 & 3 & 3 & 0 & 0 & 0 & 0 & 0 \\
ex1 & 2 & 1 & 2 & 2 & 2 & 0 & 0 & 0 & 1 & 1 & 0 & 0 & 0 & 0 & 0 \\
cnt3-5 & 13 & 0 & 0 & 0 & 0 & 19 & 0 & 0 & 0 & 0 & 0 & 0 & 0 & 0 & 0 \\
multiply\_n13 & 31 & 0 & 0 & 0 & 0 & 15 & 0 & 0 & 0 & 0 & 0 & 0 & 0 & 0 & 0 \\
rd53 & 19 & 0 & 0 & 0 & 0 & 29 & 0 & 0 & 0 & 0 & 0 & 0 & 0 & 0 & 0 \\
rd84 & 41 & 0 & 0 & 0 & 0 & 29 & 0 & 0 & 0 & 0 & 0 & 0 & 0 & 0 & 0 \\
multiplier\_n15 & 50 & 0 & 0 & 0 & 0 & 24 & 0 & 0 & 0 & 0 & 11 & 0 & 0 & 0 & 0 \\
wim & 57 & 0 & 0 & 0 & 0 & 44 & 0 & 0 & 0 & 0 & 0 & 0 & 0 & 0 & 0 \\
system9 & 92 & 0 & 0 & 0 & 0 & 69 & 0 & 0 & 0 & 0 & 0 & 0 & 0 & 0 & 0 \\
cm152a & 99 & 0 & 0 & 0 & 0 & 46 & 0 & 0 & 0 & 0 & 0 & 0 & 0 & 0 & 0 \\
sym9 & 122 & 0 & 0 & 0 & 0 & 69 & 0 & 0 & 0 & 0 & 0 & 0 & 0 & 0 & 0 \\
sym6 & 75 & 0 & 0 & 0 & 0 & 45 & 0 & 0 & 0 & 0 & 0 & 0 & 0 & 0 & 0 \\
ising\_model\_16 & 97 & 70 & 70 & 70 & 70 & 70 & 70 & 70 & 70 & 70 & 20 & 0 & 0 & 0 & 0 \\
ising\_model\_13 & 105 & 70 & 70 & 70 & 70 & 70 & 70 & 70 & 70 & 70 & 20 & 0 & 0 & 0 & 0 \\
square\_root\_n18 & 235 & 0 & 0 & 0 & 0 & 78 & 0 & 0 & 0 & 0 & 66 & 0 & 0 & 0 & 0 \\
multiplier\_n45 & 1202 & 329 & 15 & 0 & 0 & 452 & 325 & 27 & 0 & 0 & 293 & 273 & 0 & 0 & 0 \\
\bottomrule
s\_10q\_16d\_13cx & 1 & 0 & 1 & 1 & 1 & 0 & 0 & 0 & 1 & 1 & 0 & 0 & 0 & 0 & 0 \\
s\_18q\_10d\_8cx & 6 & 6 & 6 & 6 & 6 & 4 & 4 & 4 & 4 & 4 & 0 & 0 & 0 & 0 & 0 \\
s\_26q\_9d\_14cx & 8 & 8 & 8 & 8 & 8 & 6 & 6 & 6 & 6 & 6 & 0 & 0 & 0 & 0 & 0 \\
s\_26q\_9d\_15cx & 7 & 7 & 7 & 7 & 7 & 5 & 5 & 5 & 5 & 5 & 0 & 0 & 0 & 0 & 0 \\
s\_30q\_5d\_9cx & 4 & 4 & 4 & 4 & 4 & 3 & 3 & 3 & 3 & 3 & 0 & 0 & 0 & 0 & 0 \\
m\_45q\_8d\_29cx & 8 & 8 & 8 & 8 & 8 & 5 & 5 & 5 & 5 & 5 & 0 & 0 & 0 & 0 & 0 \\
m\_48q\_7d\_17cx & 9 & 9 & 9 & 9 & 9 & 5 & 5 & 5 & 5 & 5 & 0 & 0 & 0 & 0 & 0 \\
m\_53q\_18d\_48cx & 38 & 17 & 17 & 0 & 0 & 28 & 18 & 5 & 0 & 0 & 24 & 0 & 0 & 0 & 0 \\
m\_59q\_11d\_38cx & 33 & 14 & 8 & 0 & 0 & 19 & 12 & 4 & 0 & 0 & 13 & 0 & 0 & 0 & 0 \\
m\_60q\_15d\_67cx & 74 & 33 & 20 & 0 & 0 & 63 & 27 & 9 & 0 & 0 & 51 & 0 & 0 & 0 & 0 \\
l\_79q\_21d\_112cx & 182 & 112 & 51 & 34 & 28 & 69 & 32 & 21 & 13 & 0 & 56 & 24 & 0 & 0 & 0 \\
l\_83q\_21d\_129cx & 246 & 120 & 70 & 45 & 0 & 145 & 103 & 31 & 21 & 0 & 128 & 98 & 0 & 0 & 0 \\
l\_86q\_22d\_159cx & 298 & 177 & 83 & 47 & 0 & 163 & 127 & 31 & 28 & 0 & 154 & 120 & 0 & 0 & 0 \\
l\_86q\_29d\_223cx & 464 & 281 & 134 & 84 & 0 & 273 & 192 & 81 & 43 & 0 & 256 & 149 & 74 & 0 & 0 \\
l\_87q\_14d\_119cx & 184 & 109 & 57 & 22 & 0 & 108 & 54 & 23 & 18 & 0 & 97 & 22 & 0 & 0 & 0 \\
\bottomrule
\bottomrule
\end{tabular}%
\end{table*}

\begin{table*}[t]
\caption{Comprehensive Fidelity and Inter-Core Connection Counts Comparison for 10- to 90-Qubit Core Architectures. Each cell presents the result as \textbf{Fidelity (Inter-core Connection Counts)} for both low (5x) and high (100x) non-local communication latency scenarios.}
\label{tab:fidelity_appendix}
\centering

\begin{adjustbox}{width=\textwidth, center}
\begin{tabular}{@{}l|cc|cc|cc||cc|cc|cc@{}}
\toprule
\toprule
\multirow{3}{*}{\textbf{Circuit Name}} & \multicolumn{6}{c|}{\textbf{Architecture 10x10}} & \multicolumn{6}{c}{\textbf{Architecture 20x10}} \\
\cmidrule(l){2-13}
& \multicolumn{2}{c|}{\textbf{Qiskit}} & \multicolumn{2}{c|}{\textbf{QARMA}} & \multicolumn{2}{c|}{\textbf{QARMA-R}} & \multicolumn{2}{c|}{\textbf{Qiskit}} & \multicolumn{2}{c|}{\textbf{QARMA}} & \multicolumn{2}{c}{\textbf{QARMA-R}} \\
\cmidrule(l){2-3} \cmidrule(l){4-5} \cmidrule(l){6-7} \cmidrule(l){8-9} \cmidrule(l){10-11} \cmidrule(l){12-13}
& \textbf{5x} & \textbf{100x} & \textbf{5x} & \textbf{100x} & \textbf{5x} & \textbf{100x} & \textbf{5x} & \textbf{100x} & \textbf{5x} & \textbf{100x} & \textbf{5x} & \textbf{100x} \\
\midrule
\texttt{4mod5-bdd\_287} & 3.23e-02(1) & 3.36e-04(1) & 1.65e-01(0) & 1.65e-01(0) & 1.65e-01(0) & 1.65e-01(0) & 1.65e-01(0) & 1.68e-01(0) & 1.65e-01(0) & 1.65e-01(0) & 1.65e-01(0) & 1.65e-01(0) \\
\texttt{bv\_n10} & 3.35e-02(3) & 6.04e-06(3) & 7.67e-02(2) & 2.62e-04(2) & 3.12e-01(0) & 3.12e-01(0) & 3.35e-02(3) & 6.04e-06(3) & 3.00e-02(3) & 6.55e-06(3) & 3.12e-01(0) & 3.12e-01(0) \\
\texttt{cc\_n10} & 4.38e-02(3) & 6.56e-06(3) & 9.22e-02(2) & 2.63e-04(2) & 4.19e-01(0) & 4.19e-01(0) & 4.38e-02(3) & 6.56e-06(3) & 3.53e-02(3) & 5.44e-06(3) & 4.19e-01(0) & 4.19e-01(0) \\
\texttt{ex1-226} & 1.98e-01(1) & 1.81e-03(1) & 9.72e-01(0) & 9.72e-01(0) & 9.72e-01(0) & 9.72e-01(0) & 1.98e-01(1) & 1.80e-03(1) & 9.72e-01(0) & 9.72e-01(0) & 9.72e-01(0) & 9.72e-01(0) \\
\texttt{graycode6\_47} & 1.98e-01(1) & 1.81e-03(1) & 9.72e-01(0) & 9.72e-01(0) & 9.72e-01(0) & 9.72e-01(0) & 1.80e-01(1) & 1.64e-03(1) & 1.98e-01(1) & 1.54e-03(1) & 7.63e-01(0) & 7.63e-01(0) \\
\texttt{xor5\_254} & 1.98e-01(1) & 1.81e-03(1) & 9.72e-01(0) & 9.72e-01(0) & 9.72e-01(0) & 9.72e-01(0) & 1.98e-01(1) & 1.80e-03(1) & 9.72e-01(0) & 9.72e-01(0) & 9.72e-01(0) & 9.72e-01(0) \\
\midrule
\texttt{s\_10q\_16d\_13cx} & 2.16e-01(1) & 1.19e-02(1) & 5.52e-01(0) & 5.52e-01(0) & 5.52e-01(0) & 5.52e-01(0) & 5.52e-01(0) & 5.52e-01(0) & 5.52e-01(0) & 5.52e-01(0) & 5.52e-01(0) & 5.52e-01(0) \\
\texttt{s\_18q\_10d\_8cx} & 1.07e-05(6) & 2.78e-19(6) & 4.32e-04(4) & 7.43e-13(4) & 1.74e-01(0) & 1.74e-01(0) & 1.04e-05(6) & 2.73e-19(6) & 4.32e-04(4) & 7.43e-13(4) & 1.74e-01(0) & 1.74e-01(0) \\
\texttt{s\_26q\_9d\_14cx} & 5.09e-09(7) & 1.17e-31(7) & 8.37e-08(6) & 1.63e-26(6) & 4.00e-02(0) & 4.00e-02(0) & 5.09e-09(7) & 1.17e-31(7) & 8.37e-08(6) & 1.63e-26(6) & 4.00e-02(0) & 4.00e-02(0) \\
\texttt{s\_26q\_9d\_15cx} & 6.29e-08(6) & 1.16e-27(6) & 8.86e-07(5) & 1.44e-22(5) & 3.75e-02(0) & 3.75e-02(0) & 6.29e-08(6) & 1.16e-27(6) & 8.86e-07(5) & 1.44e-22(5) & 3.75e-02(0) & 3.75e-02(0) \\
\texttt{s\_30q\_5d\_9cx} & 2.25e-06(4) & 2.05e-18(4) & 4.68e-05(3) & 3.16e-16(3) & 2.02e-02(0) & 2.02e-02(0) & 2.24e-06(4) & 2.05e-18(4) & 4.68e-05(3) & 3.16e-16(3) & 2.02e-02(0) & 2.02e-02(0) \\
\bottomrule
\end{tabular}
\end{adjustbox}

\vspace{0.5cm} 

\begin{adjustbox}{width=\textwidth, center}
\begin{tabular}{@{}l|cc|cc|cc||cc|cc|cc@{}}
\toprule
\multirow{3}{*}{\textbf{Circuit Name}} & \multicolumn{6}{c|}{\textbf{Architecture 40x10}} & \multicolumn{6}{c}{\textbf{Architecture 90x10}} \\
\cmidrule(l){2-13}
& \multicolumn{2}{c|}{\textbf{Qiskit}} & \multicolumn{2}{c|}{\textbf{QARMA}} & \multicolumn{2}{c|}{\textbf{QARMA-R}} & \multicolumn{2}{c|}{\textbf{Qiskit}} & \multicolumn{2}{c|}{\textbf{QARMA}} & \multicolumn{2}{c}{\textbf{QARMA-R}} \\
\cmidrule(l){2-3} \cmidrule(l){4-5} \cmidrule(l){6-7} \cmidrule(l){8-9} \cmidrule(l){10-11} \cmidrule(l){12-13}
& \textbf{5x} & \textbf{100x} & \textbf{5x} & \textbf{100x} & \textbf{5x} & \textbf{100x} & \textbf{5x} & \textbf{100x} & \textbf{5x} & \textbf{100x} & \textbf{5x} & \textbf{100x} \\
\midrule
\texttt{4mod5-bdd\_287} & 1.65e-01(0) & 1.68e-01(0) & 1.65e-01(0) & 1.65e-01(0) & 1.65e-01(0) & 1.65e-01(0) & 1.65e-01(0) & 1.68e-01(0) & 1.65e-01(0) & 1.65e-01(0) & 1.65e-01(0) & 1.65e-01(0) \\
\texttt{bv\_n10} & 3.35e-02(3) & 6.04e-06(3) & 3.00e-02(3) & 6.55e-06(3) & 3.12e-01(0) & 3.12e-01(0) & 3.35e-02(3) & 6.04e-06(3) & 3.00e-02(3) & 6.55e-06(3) & 3.12e-01(0) & 3.12e-01(0) \\
\texttt{cc\_n10} & 4.38e-02(3) & 6.56e-06(3) & 3.53e-02(3) & 5.44e-06(3) & 4.19e-01(0) & 4.19e-01(0) & 4.38e-02(3) & 6.56e-06(3) & 3.53e-02(3) & 5.44e-06(3) & 4.19e-01(0) & 4.19e-01(0) \\
\texttt{ex1-226} & 1.98e-01(1) & 1.80e-03(1) & 9.72e-01(0) & 9.72e-01(0) & 9.72e-01(0) & 9.72e-01(0) & 1.98e-01(1) & 1.80e-03(1) & 1.98e-01(1) & 1.54e-03(1) & 8.12e-01(0) & 8.12e-01(0) \\
\texttt{graycode6\_47} & 1.89e-01(1) & 1.72e-03(1) & 1.98e-01(1) & 1.54e-03(1) & 7.63e-01(0) & 7.63e-01(0) & 1.89e-01(1) & 1.72e-03(1) & 1.98e-01(1) & 1.54e-03(1) & 7.63e-01(0) & 7.63e-01(0) \\
\texttt{xor5\_254} & 1.98e-01(1) & 1.80e-03(1) & 9.72e-01(0) & 9.72e-01(0) & 9.72e-01(0) & 9.72e-01(0) & 1.98e-01(1) & 1.80e-03(1) & 1.98e-01(1) & 1.54e-03(1) & 8.12e-01(0) & 8.12e-01(0) \\
\midrule
\texttt{s\_10q\_16d\_13cx} & 2.20e-01(1) & 1.21e-02(1) & 5.52e-01(0) & 5.52e-01(0) & 5.52e-01(0) & 5.52e-01(0) & 2.23e-01(1) & 1.22e-02(1) & 2.09e-01(1) & 1.20e-02(1) & 4.44e-01(0) & 4.44e-01(0) \\
\texttt{s\_18q\_10d\_8cx} & 1.10e-05(6) & 2.87e-19(6) & 4.32e-04(4) & 7.43e-13(4) & 1.74e-01(0) & 1.74e-01(0) & 1.04e-05(6) & 2.73e-19(6) & 4.32e-04(4) & 7.43e-13(4) & 1.74e-01(0) & 1.74e-01(0) \\
\texttt{s\_26q\_9d\_14cx} & 5.09e-09(7) & 1.17e-31(7) & 8.37e-08(6) & 1.63e-26(6) & 4.00e-02(0) & 4.00e-02(0) & 5.09e-09(7) & 1.17e-31(7) & 8.37e-08(6) & 1.63e-26(6) & 4.00e-02(0) & 4.00e-02(0) \\
\texttt{s\_26q\_9d\_15cx} & 6.29e-08(6) & 1.16e-27(6) & 8.86e-07(5) & 1.44e-22(5) & 3.75e-02(0) & 3.75e-02(0) & 6.29e-08(6) & 1.16e-27(6) & 8.86e-07(5) & 1.44e-22(5) & 3.75e-02(0) & 3.75e-02(0) \\
\texttt{s\_30q\_5d\_9cx} & 2.24e-06(4) & 2.05e-18(4) & 4.68e-05(3) & 3.16e-16(3) & 2.02e-02(0) & 2.02e-02(0) & 2.24e-06(4) & 2.05e-18(4) & 4.68e-05(3) & 3.16e-16(3) & 2.02e-02(0) & 2.02e-02(0) \\
\bottomrule
\bottomrule
\end{tabular}
\end{adjustbox}
\end{table*}

\begin{figure}[t]
    \centering
    \includegraphics[width=0.95\linewidth]{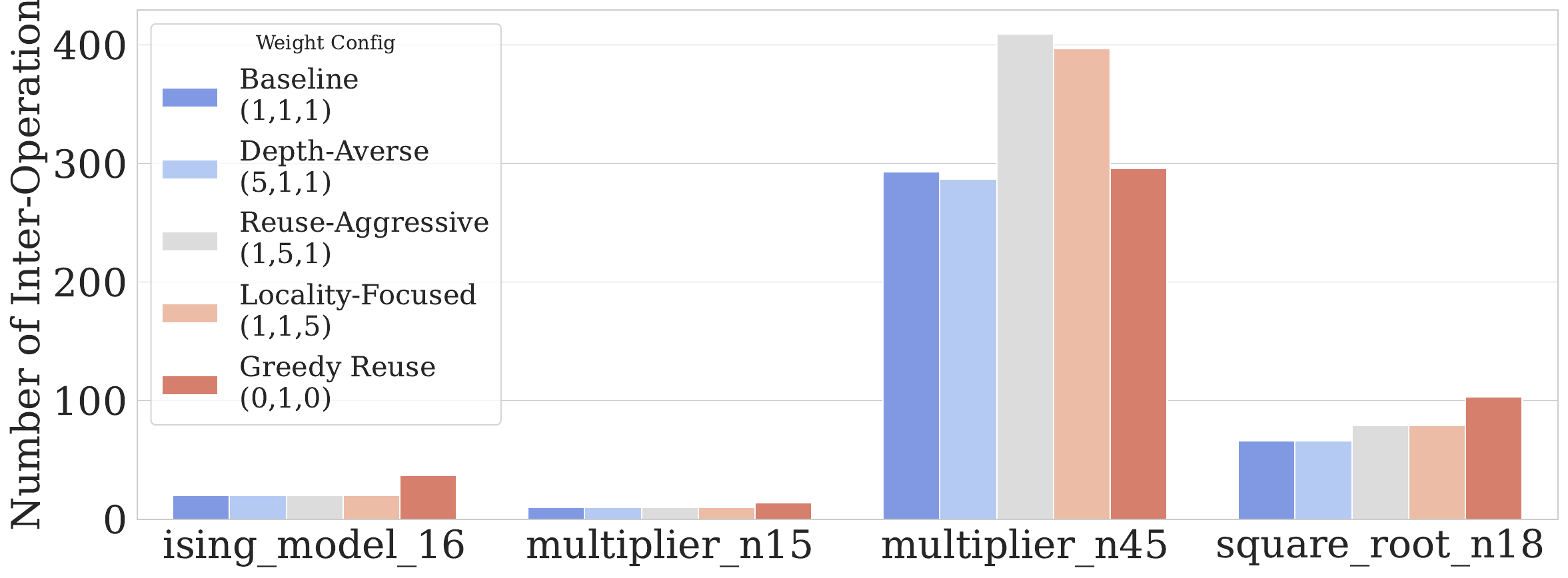}
    \caption{Sensitivity analysis of qubit reuse weights ($w_0, w_1, w_2$). Comparison of inter-operation counts on a selected $2\times5$ grid with 10 qubits per core for a common practical architecture.}
    \label{fig:sensitivity}
\end{figure}

\section{Impact of Weight Configurations on Inter-Core Communication}
\label{sec:sensitivity_analysis}

In Section~\ref{subsec:qubit_reuse_impl}, we defined a weighted cost function (Eq.~\ref{eq:qubit_reuse_weight}) to select optimal qubit reuse pairs, balancing the increase in circuit depth ($w_0$), the earliness of the reuse opportunity ($w_1$), and the duration of the gap between operations ($w_2$). In our main evaluation, we employed a balanced configuration ($w_0=w_1=w_2=1$). To assess the robustness of our approach and the impact of these hyperparameters, we conducted a sensitivity analysis on a subset of complex benchmark circuits using five distinct weight configurations (see Table~\ref{tab:sensitivity_weights}).

Our analysis reveals that for low- and medium-complexity, standard benchmarks (e.g., \texttt{bv-n10}, \texttt{4mod5-bdd}, and \texttt{system9}), the choice of weights has a negligible impact, as the reuse opportunities are straightforward and the optimal solution is stable, as shown in Table~\ref{tab:sensitivity_full} (All weight configurations resulted in zero inter-core operations). However, for large-scale, complex circuits, the configuration significantly influences the inter-core communication overhead.

As shown in Fig.~\ref{fig:sensitivity}, we observed that the \textit{Depth-Averse} strategy ($5, 1, 1$) yielded better performance. For the complex \texttt{multiplier-n45} circuit, increasing the penalty on circuit depth reduced the inter-core connection count to 287, slightly outperforming the Baseline (293 connections). This indicates that maintaining a compact temporal schedule (low depth) implicitly aids the spatial mapper by preventing the logical circuit from stretching excessively, which would otherwise force qubits onto distant cores.

Conversely, strategies that aggressively prioritize the ``earliness'' of reuse ($1, 5, 1$) or ``locality'' of the gap ($1, 1, 5$) proved counter-productive for complex topologies. The \textit{Reuse-Aggressive} configuration caused a sharp degradation in performance, increasing the inter-core connections for \texttt{multiplier-n45} to 409, a 39\% increase over the Baseline. Similarly, the \textit{Greedy} approach ($0, 1, 0$), which ignores depth costs, resulted in high variance; while it performed adequately on some circuits, it doubled the inter-core connections for \texttt{ising-model-16} (37 connections) compared to the Baseline (20 connections).

These results suggest that while the balanced Baseline configuration ($1, 1, 1$) provides robust performance across a wide range of circuits, the primary driver for minimizing inter-core communication in modular architectures is reducing circuit depth. Therefore, future optimizations should favor Depth-Averse weighting to prevent the formation of deep dependency chains that complicate the spatial allocation of qubits.

\section{Dynamic Action Mask Generation}
\label{app:action_mask}

To ensure the validity of the qubit allocation solutions generated by the QARMA policy, we employ a dynamic action masking mechanism during autoregressive decoding. This mechanism restricts the agent's action space $\mathcal{A}_t$ at each decoding step $t$ to only those physical cores that satisfy the architectural hard constraints.

Let $l_{t} \in \mathbb{R}^{|\mathcal{C}|}$ be the raw logits output by the policy network for the current logical qubit $q_t$, where $|\mathcal{C}|$ is the number of cores. We apply a binary mask $M_t \in \{0, -\infty\}^{|\mathcal{C}|}$ before the softmax operation:

\begin{equation}
    P(a_t | s_t) = \text{softmax}(l_{t} + M_t)
\end{equation}

The mask $M_t$ is updated dynamically at each step based on the current state of the environment. For a target core $c$, the mask value $M_t[c]$ is defined as:

\begin{equation}
    M_t[c] = 
    \begin{cases} 
      0 & \text{if } \text{Valid}(q_t, c) \\
      -\infty & \text{otherwise}
    \end{cases}
\end{equation}

The validity function $\text{Valid}(q_t, c)$ serves as the conjunction of two primary constraints:

\textbf{1. Capacity Constraint:} The core $c$ must have available physical qubits. Let $U_t[c]$ be the current qubit usage of core $c$ and $C_{max}$ be its capacity.
\begin{equation}
    \text{Cond}_{cap}(c) \iff U_t[c] < C_{max}
\end{equation}
Note that if $q_t$ is a reused qubit (i.e., it reuses a physical slot freed by a measured qubit), it effectively does not increment $U_t[c]$, preventing false capacity blocks.

\textbf{2. Topology Constraint:} If the logical qubit $q_t$ has a strict dependency on a previously placed qubit $q_{prev}$ (e.g., a two-qubit gate in the current slice), the target core $c$ must be reachable from the core assigned to $q_{prev}$. For architectures with restricted connectivity graphs $G=(V, E)$:
\begin{equation}
    \text{Cond}_{topo}(c) \iff (c, \text{loc}(q_{prev})) \in E \lor c = \text{loc}(q_{prev})
\end{equation}
In our experiments, where inter-core transfers are costly but fully connected (all-to-all modular grid), this constraint is relaxed to allow any transfer, relying on the reward signal to punish distance. However, for strictly sparse topologies, this mask forces valid routing paths.

The final mask is the intersection of these conditions:
\begin{equation}
    \text{Valid}(q_t, c) = \text{Cond}_{cap}(c) \land \text{Cond}_{topo}(c)
\end{equation}

By setting the logits of invalid actions to $-\infty$, the probability of selecting an illegal core becomes exactly zero, guaranteeing that the DRL agent only samples valid allocation strategies.

\section{Detailed Inter-Core Connections}
\label{appendix:detail_inter-core_benchmark}
This appendix provides a detailed breakdown of the primary optimization metric used in this study: the number of inter-core connections. Table~\ref{tab:icc_bench_circuits} presents the complete results for all benchmark circuits across the five tested modular architectures, which scale from 10 to 90 qubits per core.

\section{Detailed Fidelity Benchmark Results}
\label{appendix:detail_fidelity_benchmark}

This appendix provides the comprehensive data from our fidelity analysis experiments. While Fig.~\ref{fig:fidelity_standard_and_random}a and Fig.~\ref{fig:fidelity_standard_and_random}b in the main text visualize the full sensitivity analysis across all tested latency factors (5x, 10x, 20x, 40x, 60x, 80x, and 100x), presenting this entire dataset in tabular form would be excessive and consume a great deal of space.

Therefore, to clearly illustrate the impact of communication overhead while maintaining readability, Table~\ref{tab:fidelity_appendix} presents the results for two distinct non-local communication latency scenarios: a low-latency case (5x) and a high-latency case (100x). These two scenarios were chosen because they effectively represent the boundary conditions of our analysis, making the performance distinctions between the different compilation methods most apparent without including redundant data from intermediate latency points.

%

\end{document}